\documentclass[11pt]{article}
\usepackage{mydef2col}
\usepackage{markArticle}

\title{DSINC beyond Heston: the 3/2 and other non-affine\\ stochastic-volatility models}

\shorttitle{DSINC beyond Heston}

\author{
    \authorstyle{ Andrey Itkin}
    \newline \newline
    \institution{FRE Department, Tandon School of Engineering, New York University, email: \url{aitkin@nyu.edu}} \\
}

\date{\today}

\begin{document}

\maketitle

\lettrineabstract{This paper extends the damped-sinc (DSINC) method for American options beyond the affine Heston class to general one-factor stochastic-volatility models. We show that the collapse identity, which reduces the early-exercise integral to a single spectral integral, remains valid for these models. Damping between zero and one is admissible whenever the discounted spot is a nonnegative local martingale. An Itô transformation of the volatility factorizes the conditional characteristic function and identifies a natural coordinate for approximating the exercise boundary. The 3/2 model admits closed-form joint transforms in terms of Kummer functions, while uncorrelated SABR admits a Bessel representation. When a practical closed-form transform is unavailable, the required volatility integral is computed directly by solving one backward equation per frequency, without a grid in the log-price. Through a known change of volatility coordinate, this approach also accommodates correlated SABR. Tests on the 3/2, SABR, GARCH and lognormal-volatility models show agreement with fine finite-difference benchmarks to within one basis point of the strike. Computational cost is comparable to or lower than that of our finite-difference implementation, except for the slower Hankel implementation of uncorrelated SABR.
}

%

\section{Introduction} \label{sec:intro}

The damped-sinc (DSINC) method of \cite{AndersenItkinKazbek2026} was introduced to price American options and flexible forward contracts on exchange rates via Volterra integral equations in time-inhomogeneous stochastic-volatility models (SVM). A particular implementation was presented for the Heston model. The central result of the method is a \emph{collapse identity}. Under band-limitation hypotheses\footnote{Band-limitation means that the Fourier transform of the exponentially damped transition density vanishes outside the frequency band resolved by the sinc grid, so that sinc interpolation of the density is exact by the Shannon sampling theorem. For actual densities this holds only approximately, with an explicitly bounded truncation error.}, the sum over a local DSINC basis reduces to the conditional characteristic function (CF) of the log-price on a reflected contour\footnote{For a real frequency $\omega$ and damping exponent $\chi$, the CF is evaluated at $-\omega-\iu\chi$, that is, on the real axis reversed in sign and shifted by the damping to the line with imaginary part $-\chi$.}.

The log-price integral over the exercise region then becomes a single spectral pairing\footnote{By a spectral pairing we mean the integral of the payoff against the transition density, evaluated in the Fourier domain by Parseval's identity.} with an explicit payoff transform. The identity follows from the spectral representation and does not use affine dynamics. The affine structure is used only afterwards, when the remaining variance integral is evaluated through tilted variance transforms and a M\"obius propagator.

There are good reasons to consider models outside the affine class. Studies that combine returns, realized variance and option prices for the S\&P~500 favor variance dynamics in which the volatility of variance grows faster than the square root of the variance, \cite{ChristoffersenJacobsMimouni2010, KaeckAlexander2012}. The 3/2 SVM has this property and keeps a closed-form CF, \cite{Lewis2000, CarrSun2007}. It is used for modeling realized-variance and VIX derivatives in \cite{Drimus2012, BaldeauxBadran2014}, and the 4/2 model of \cite{Grasselli2017} combines it with the Heston model. The GARCH diffusion of \cite{Nelson1990} has a variance whose diffusion coefficient is proportional to its level, and no closed-form transform, \cite{BaroneAdesiEtAl2005}. The SABR model of \cite{Hagan2002}, the market standard for interpolating smiles, has a lognormal volatility. Other solvable non-affine specifications appear in \cite{Itkin2013}.

American options under stochastic volatility are usually priced by finite differences (FD), with operator splitting or penalty methods for the early-exercise constraint, \cite{ClarkeParrott1999, IkonenToivanen2009, HaentjensInTHout2015, Andersen2025}. Fourier methods for early exercise use cosine expansions, \cite{FangOosterlee2011}, Shannon wavelets, \cite{MareeOrtizGraciaOosterlee2017} or sinc expansions, \cite{BaschettiEtAl2022}. They need a CF, which is available in closed form mainly for affine and L\'evy models. Outside the affine class, two features change. The CF of the log-price, conditional on the terminal volatility, loses its exponential-affine form. In addition, exponential moments of the log-price can be infinite, \cite{AndersenPiterbarg2007,LionsMusiela2007}, which restricts the admissible damping. Closely related integral-equation and CF treatments of early exercise, outside the affine class, appear in \cite{ItkinCF2025j} which also provides a survey of the corresponding literature.

This paper extends DSINC to such models. First, we restate the collapse theorem for any one-factor SVM and add a second damping range, the strip between zero and one, at the cost of one residue term. This strip is admissible whenever the discounted spot is a nonnegative local martingale. In contrast, the Put damping of \cite{AndersenItkinKazbek2026} is never admissible in SABR, where the forward can be absorbed at zero, or in models with lognormal volatility such as the GARCH diffusion.

Second, an It\^o transformation of the volatility factor splits the conditional CF into an explicit endpoint phase and a bridge functional of a complex potential. The same transformation defines the coordinate in which the exercise boundary should be approximated as piecewise linear, and the transform needed for the volatility integral solves one backward equation per frequency.

Third, in the 3/2 model this transform is a Kummer function. This gives the full-range volatility integral in closed form and explicit conditions on the damping.

Fourth, for SABR with any elasticity $\beta$ and zero correlation $\rho$, we combine the Bessel representation of \cite{ItkinMuravey2021SABR} with the spectral pairing. The premium integrand over the exercise region then has a closed form in the Hankel representation, and each Bessel frequency requires one equation in the volatility. In contrast, nonzero correlation couples the Bessel frequencies. We also outline a perturbation expansion around the uncorrelated case. Taking the instantaneous lognormal volatility of the forward as the volatility factor instead removes the forward from the coefficients. This symmetry reduction was used by \cite{Park2013} for European options, and with it correlated SABR is solved like the general models below. The semi-analytic program for American options in time-dependent one-factor models, of which this method is a part, is surveyed in \cite{ItkinMuravey2024jd}.

Fifth, for general models we solve the fixed-frequency equation numerically and sample the exercise boundary at the volatility nodes, which removes the band-mass approximation\footnote{In \cite{AndersenItkinKazbek2026}, the volatility integral is assembled from the contributions of the individual segments of the piecewise-linear approximation, weighted by the probabilities of the corresponding volatility segments. This band-mass approximation has an error distinct from that of the collapse identity, and in general refining the segments alone does not remove it.}. Tests on the 3/2, SABR, GARCH and lognormal-volatility models confirm the closed forms, the damping strip and the American prices.

The rest of the paper is organized as follows. \Cref{sec:theory} extends the collapse theorem to general one-factor SVMs, adds the second damping range, and derives the It\^o factorization together with the fixed-frequency backward equation. \Cref{sec:32} treats the 3/2 model, where the volatility integral is available in closed form through Kummer functions, and states the admissible damping conditions. \Cref{sec:sabr} develops the Bessel representation for SABR with zero correlation, the regularized pairing with the mass at zero and the correlation expansion, and reports numerical tests for zero correlation. \Cref{sec:general} describes the fixed-frequency numerical solution for general models, the nodal boundary representation and the time recursion, and presents the numerical experiments for the 3/2, lognormal-volatility and GARCH specifications. \Cref{sec:concl} summarizes what carries over from the Heston model and concludes.

\section{Collapse beyond affine dynamics} \label{sec:theory}

Our notation hereafter follows \cite{AndersenItkinKazbek2026}. Let $S$ be the exchange rate and $K$ the strike. The log-moneyness at a date $u$ is $x_u = \log(S_u/K)$. The domestic and foreign rates $r_d$ and $r_f$ are constant. Deterministic rates change only the notation below. The volatility factor $y$ takes values in an open interval $\mathcal{Y}\subseteq(0,\infty)$ and enters the log-price through the instantaneous variance $V(y)>0$. Under the domestic risk-neutral measure,
\begin{equation} \label{eq:gen}
dx_u  =  \bigl(r_d - r_f - \tfrac12 V(y_u)\bigr)\,du  +  \sqrt{V(y_u)}\,dW^S_u, \qquad dy_u  =  \mu(y_u)\,du  +  \eta(y_u)\,dW^y_u,
\end{equation}
where $W^S$ and $W^y$ are Brownian motions with correlation $|\rho|<1$, $\mu$ is the drift of the volatility factor and $\eta>0$ is its diffusion coefficient (vol-of-vol). We assume sufficient local regularity for a non-explosive solution and for the It\^o transformations below, and that $y$ stays in $\mathcal{Y}$. The coefficients in \eqref{eq:gen} do not depend on $x$. This translation invariance gives the independent fixed-frequency equations, while the collapse identity itself only requires a suitable transition law. A local stochastic volatility coefficient depending on both $x$ and $y$ generally couples Fourier frequencies. SABR, whose volatility coefficient depends on the price, is treated separately.

Four models fit \eqref{eq:gen}. Three of them take $y = v$, the instantaneous variance, with $V(v) = v$, $W^v = W^y$, mean-reversion rate $\kappa$, level $\theta$ and volatility of variance $\xi$. The Heston model has $dv  =  \kappa( \theta - v)\,dt  +  \xi\sqrt{v}\,dW^v$. The 3/2 model has $dv = \kappa v(\theta-v)\,dt  +  \xi v^{3/2}\, dW^v$, so its rate of mean reversion is $\kappa v$. The GARCH diffusion of \cite{Nelson1990} has $dv = \kappa(\theta-v)\,dt + \xi v\,dW^v$. The lognormal-volatility model takes $y = \sigma$, the volatility itself, with $V(\sigma)  =  \sigma^2$, $\mu = 0$ and $\eta(\sigma)  =  \varepsilon \sigma$. It coincides with the SABR model with unit elasticity and with the model of \cite{HullWhite1987} with a driftless volatility. We write $\varepsilon$ for its volatility of volatility because $\nu$ denotes Bessel orders below.

For $t<u$, let $p(x_u,y_u\mid x,y)$ be the joint transition density of $(x_u,y_u)$ given $x_t = x$ and $y_t = y$, and let $p_y(y_u\mid y)$ be the density of $y_u$ given $y_t = y$. The integral-equation approach writes the value of an American option as a European component plus an early-exercise premium (EEP), with value matching determining the exercise boundary through a nonlinear Volterra equation, see \cite{ItkinCF2025j} and references therein. Related constructions exist for the constant-parameter Heston model, \cite{tzavalis2003pricing, Chiarella2004, Chiarella2005}, jump-diffusions, \cite{Chiarella2009,Itkin2024jd,ItkinCF2025j} and the 3/2 model, \cite{DetempleKitapbayev2017, ItkinCF2025j}. For general diffusions the decomposition rests on the change-of-variables formula with local time on surfaces of \cite{Peskir2005, Peskir2007}. Building on these results, \cite{ItkinKitapbayev2025r,ItkinCF2025j} develop early-exercise decompositions in general time-inhomogeneous settings, in which the premium integrand is determined by the generator and the payoff. Under their regularity and boundary hypotheses, the decomposition for the model \eqref{eq:gen} and an American Put with expiry $T$ reads
\begin{equation} \label{eq:eep}
P(t,x,y)  =  P_E(t,x,y)  +  K\int_t^T e^{-r_d(u-t)}\int_{\mathcal{Y}} J(u,y_u ; x,y)\,dy_u\,du,
\end{equation}
where
\begin{equation} \label{eq:J}
J(u,y_u ; x,y)  =  \int_{-\infty}^{x^*(u,y_u)} g(x_u)\,p(x_u,y_u\mid x,y)\,dx_u, \qquad g(x) = r_d-r_f e^{x}.
\end{equation}
Here $P$ and $P_E$ are the American and European Put prices, $x^*(u,y)$ is the exercise boundary in log-moneyness and $g$ is the payoff weight. We refer to $J$ as the premium integrand. The boundary is determined through the value-matching condition $P(t,x^*(t,y),y) = K\bigl(1-e^{x^*(t,y)}\bigr)$.

\begin{myremark}[Martingale property] \label{rem:martingale}
The derivation of \eqref{eq:eep} uses only that the Put value and the payoff weight are bounded on the exercise region. The local martingale that appears in it is then bounded, hence a true martingale. So \eqref{eq:eep} holds even when the discounted spot $e^{-(r_d-r_f)t}S_t$ is a strict local martingale. The delayed-exercise premium (DEP) form and the Put--Call duality of \cite{AndersenItkinKazbek2026} need a true martingale. The condition for this is model specific.
\end{myremark}

\subsection{The collapse theorem on two damping ranges} \label{sec:collapse}

To evaluate the premium integrand $J$ in \eqref{eq:J}, the DSINC method represents the joint transition density as a function of the terminal log-moneyness, with the terminal volatility held fixed. We first introduce the interpolation basis and the exponential weighting used in this representation.

Let $\Delta x>0$ be the grid spacing and $x_j=j\Delta x$, $j\in\mathbb{Z}$, the grid nodes. The normalized sinc function is
\begin{equation*}
\operatorname{sinc}(s)=\frac{\sin(\pi s)}{\pi s}, \qquad s\ne0, \qquad \operatorname{sinc}(0)=1.
\end{equation*}
Thus, $\operatorname{sinc}((x_u-x_j)/\Delta x)$ equals one at $x_u=x_j$ and zero at every other grid node. For a real parameter $\chi$, called the damping exponent, the damped-sinc basis is
\begin{equation*}
\phi_j(x_u)=e^{-\chi(x_u-x_j)}\operatorname{sinc}\bigl((x_u-x_j)/\Delta x\bigr).
\end{equation*}
The exponential factor preserves these interpolation values. Expanding the density in this basis is equivalent to applying ordinary sinc interpolation to the exponentially weighted density
\begin{equation*}
p_\chi(x_u)=e^{\chi x_u}p(x_u,y_u\mid x,y).
\end{equation*}
The choice of $\chi$ controls the integrability of this weighted density and of the corresponding payoff terms; the two admissible damping ranges are specified below.

The frequency band associated with the grid is $\Omega=[-\pi/\Delta x,\pi/\Delta x]$, called the Nyquist band. If $p_\chi$ is band-limited to $\Omega$, meaning that its Fourier transform vanishes outside this interval, sinc interpolation gives the exact representation
\begin{equation*}
p(x_u,y_u\mid x,y)=\sum_{j\in\mathbb{Z}}c_j\phi_j(x_u), \qquad c_j=p(x_j,y_u\mid x,y).
\end{equation*}
The dependence of $c_j$ on the dates and conditioning state is suppressed here. The treatment of densities that are not band-limited is given in \cref{rem:bandlimit}.

For a real Fourier frequency $\omega$, the exponential weighting shifts the transform to a complex argument. In the Fourier pairing used below, this argument is $\widetilde\omega=-\omega-\iu\chi$, which we call the reflected frequency: the minus sign comes from the pairing, and the imaginary shift comes from the damping. The model enters through the transform of the joint density in the log-price,
\begin{equation} \label{eq:Psi}
\Psi(\omega,y_u ; x,y) = \int_{-\infty}^{\infty}e^{\iu\omega x_u}\,p(x_u,y_u\mid x,y)\,dx_u.
\end{equation}
Divided by $p_y$, it becomes the CF of $x_u$ conditional on the terminal volatility,
\begin{equation} \label{eq:condCF}
\psi(\omega,y_u ; x,y) = \frac{\Psi(\omega,y_u ; x,y)}{p_y(y_u\mid y)} = \EE\bigl[ e^{\iu\omega x_u} \mid y_u,\ x_t  =  x,\ y_t  =  y \bigr].
\end{equation}
Let $\mathcal{K} = x^*(u,y_u)$ be the exercise level. The damped payoff transform is
\begin{equation} \label{eq:Ptilde}
\widetilde P(\omega,y_u) =  e^{(\iu\omega - \chi) \mathcal{K}} \Bigl[ \frac{r_d}{\iu\omega - \chi} - \frac{r_f\,e^{\mathcal{K}}}{1 + \iu\omega-\chi}\Bigr].
\end{equation}
For $\chi<0$, \eqref{eq:Ptilde} is the convergent transform $\int_{ -\infty}^ {\mathcal{K}} g(x) e^{(\iu\omega - \chi)x}\,dx$. For $0<\chi<1$, this integral diverges when $r_d\ne0$. The algebraic expression in \eqref{eq:Ptilde} nevertheless remains finite and represents the transform of the regularized payoff weight
\begin{equation*}
\widetilde P(\omega,y_u)
= \int_{\mathbb{R}}\bigl[ g(x) \mathbf{1}_{ \{x\le\mathcal{K}\}} - r_d \bigr]
e^{(\iu\omega-\chi)x}\,dx.
\end{equation*}
The subtraction removes the nonzero limit of the payoff weight as $x\to-\infty$, while the exponential factor ensures integrability as $x\to+\infty$. Restoring the subtracted constant produces the additional term $r_d p_y(y_u\mid y)$ in \eqref{eq:parseval}.

\begin{theorem}[Collapse and Parseval reduction for one-factor models] \label{thm:collapse}
Fix $t<u$ and a state $(x,y)$. The assertions below are understood for almost every terminal volatility $y_u\in\mathcal{Y}$. Assume the following:
\begin{enumerate}
\item[\normalfont(H0)] The moment $\EE[e^{\chi x_u}\mid x_t = x,\ y_t = y]$ is finite.
\item[\normalfont(H1)] The damped density $p_\chi(x_u) = e^{\chi x_u}p(x_u,y_u\mid x,y)$ is band-limited to $\Omega$.
\item[\normalfont(H2)] Either $\chi<0$ or $0<\chi<1$.
\end{enumerate}
Then the following hold.

\emph{(i) Collapse identity.} On the band,
\begin{equation} \label{eq:collapse}
\sum_j c_j(u,y_u\mid x,y)\,e^{-(\iu\omega-\chi)x_j} = \frac{1}{\Delta x}\,\Psi(\widetilde\omega,y_u ; x,y), \qquad \omega\in\Omega.
\end{equation}

\emph{(ii) Parseval reduction.} The premium integrand in \eqref{eq:J} equals
\begin{equation} \label{eq:parseval}
J(u,y_u ; x,y) = \mathbf{1}_{\{\chi>0\}}\,r_d\,p_y(y_u\mid y) + \frac{1}{2\pi}\int_\Omega\Psi(\widetilde\omega,y_u ; x,y)\,\widetilde P(\omega,y_u)\,d\omega.
\end{equation}

\emph{(iii) Admissibility of the fundamental strip.} If $0<\chi<1$ (a range we call the fundamental strip) and $e^{-(r_d-r_f)s}S_s$ is a nonnegative local martingale, then (H0) holds.
\end{theorem}

\begin{proof}
By (H0) the transform $\hat p_\chi(\omega) = \int p_\chi(x_u)e^{\iu\omega x_u}dx_u = \Psi(\omega-\iu\chi,y_u ; x,y)$ exists for almost every $y_u$. Part (i) and part (ii) for $\chi<0$ are the statement of \cite{AndersenItkinKazbek2026}. Its proof uses only (H1), the sign of $\chi$ and the existence of $\hat p_\chi$. It does not use the form of $\Psi$. Let now $0<\chi<1$. Part (i) is unchanged. For part (ii), split the premium integrand as
\begin{equation*}
J = r_d\int_{-\infty}^{\infty}p\,dx_u-r_d\int_{\mathcal{K}}^{\infty}p\,dx_u-r_f\int_{-\infty}^{\mathcal{K}}e^{x_u}p\,dx_u.
\end{equation*}
The first integral is $p_y(y_u\mid y)$. The second and third are pairings $\int p_\chi h\,dx_u$ with $h_1(x) = e^{-\chi x}\mathbf{1}_{\{x>\mathcal{K}\}}$ and $h_2(x) = e^{(1-\chi)x}\mathbf{1}_{\{x\le\mathcal{K}\}}$, which are integrable because $0<\chi<1$. Parseval's identity gives $\int p_\chi h\,dx_u = (2\pi)^{-1}\int\hat p_\chi(-\omega)\hat h(\omega)\,d\omega$, with $\hat p_\chi(-\omega) = \Psi(\widetilde\omega,y_u ; x,y)$ and
\[
\hat h_1(\omega) = -e^{(\iu\omega-\chi)\mathcal{K}}/(\iu\omega-\chi), \qquad \hat h_2(\omega) = e^{(1 + \iu\omega-\chi)\mathcal{K}}/(1 + \iu\omega-\chi).
\]
The two pairings combine into \eqref{eq:Ptilde}, and (H1) restricts the integral to $\Omega$. For (iii), a nonnegative local martingale is a supermartingale, so $\EE[S_u]\le S_te^{(r_d-r_f)(u-t)}$. Jensen's inequality for the concave map $s\mapsto s^\chi$ then gives $\EE[e^{\chi x_u}]\le\bigl(S_te^{(r_d-r_f)(u-t)}/K\bigr)^{\chi}<\infty$.
\end{proof}

The identity is a property of damped-sinc interpolation. When the damped density is band-limited, the basis sum reproduces its transform. The Parseval reduction needs only the payoff transform $\widetilde P$, which depends on the payoff weight $g$ and the exercise level, not on the model. The model enters through $\Psi$ alone.

\begin{myremark}[Band-limitation] \label{rem:bandlimit}
A transition density is never exactly band-limited, so (H1) is an idealization, and it can be dropped. Without it, the proof of part (ii) gives \eqref{eq:parseval} with $\Omega$ replaced by $\mathbb{R}$, provided $p_\chi$ is square integrable and the spectral integral converges absolutely. The right-hand side of \eqref{eq:parseval} then differs from $J$ by at most
\begin{equation} \label{eq:tail}
E_{\mathrm{tail}}  =  \frac{1}{2 \pi} \int_{\mathbb{R} \setminus \Omega} \bigl| \Psi( \widetilde\omega,y_u;x,y)\, \widetilde P(\omega,y_u)\bigr|\, d\omega.
\end{equation}
Part (i) holds exactly for the projection of $p_\chi$ onto $\Omega$, the band-limited function whose transform equals $\hat p_\chi$ on $\Omega$ and vanishes outside it, with $c_j$ equal to $e^{-\chi x_j}$ times its sample at $x_j$. The band then controls an explicit truncation error instead of entering as an assumption.
\end{myremark}

\begin{myremark}[The residue] \label{rem:residue}
As a function of $\widetilde\omega$, the payoff transform has simple poles at $\widetilde\omega = 0$ and $\widetilde\omega = -\iu$, because $\iu\omega-\chi = -\iu\widetilde\omega$ and $1 + \iu\omega-\chi = 1-\iu\widetilde\omega$. The integration line $\operatorname{Im}\widetilde\omega = -\chi$ lies above both poles when $\chi<0$ and between them when $0<\chi<1$. Moving the line across the pole at zero picks up the residue $r_d\Psi(0,y_u ; x,y) = r_d\,p_y(y_u\mid y)$, which is the extra term in \eqref{eq:parseval}. The integration contour of the option formula of \cite{Lewis2001} lies in the same strip. In models where the price can be absorbed at zero, such as SABR, the law of $x_u$ has an atom at $-\infty$. On the fundamental strip \eqref{eq:parseval} still holds, with $p_y$ the full density of $y_u$. The atom enters through the residue term with weight $r_d$, and it contributes nothing to $\Psi(\widetilde\omega,y_u ; x,y)$, because $e^{\chi x}$ vanishes at $x = -\infty$. Negative damping is then unavailable, because every negative moment of the spot is infinite.
\end{myremark}

\begin{myremark}[Choice of damping] \label{rem:chichoice}
When both ranges are admissible, \eqref{eq:parseval} gives the same value for every admissible $\chi$, so a plateau in $\chi$ remains a useful diagnostic. We use $\chi = 1/2$ for non-affine models. On the fundamental strip the residue term integrates to $r_d$ over $\mathcal{Y}$, and to $r_d$ times the probability of a segment over that segment, so it adds no computational cost. The spectral term then carries the complement of the exercise region. It cancels part of the residue when the exercise probability is small, which reduces the relative accuracy far from the money.
\end{myremark}

\subsection{It\^o factorization} \label{sec:ito}

The Heston structure of \cite{AndersenItkinKazbek2026} rests on an application of It\^o's formula. Conditioning on the terminal variance is enough, because the stochastic integral of the volatility against $W^v$ can be written through the endpoints and a time integral. The same argument applies to every model of the form \eqref{eq:gen}. Define the It\^o coordinate $F$ and its drift $\ell$ by
\begin{equation} \label{eq:itocoord}
F'(y) = \frac{\sqrt{V(y)}}{\eta(y)}, \qquad \ell(y) = \mu(y)F'(y) + \tfrac12\eta(y)^2F''(y),
\end{equation}
so that $dF(y_s) = \ell(y_s)\,ds + \sqrt{V(y_s)}\,dW^y_s$ by It\^o's formula. The It\^o coordinate has the same instantaneous variance as the log-price, $d\langle F(y)\rangle_s = d\langle x\rangle_s = V(y_s)\,ds$. For a frequency $\omega$, define the complex potential
\begin{equation} \label{eq:potential}
\Upsilon_\omega(y) = \tfrac12\omega^2(1-\rho^2)V(y) + \iu\omega\bigl[\tfrac12V(y) + \rho\,\ell(y)\bigr]
\end{equation}
and the bridge functional
\begin{equation} \label{eq:bridgefun}
H_\omega(y,y_u) = \EE\Bigl[\exp\Bigl(-\int_t^u\Upsilon_\omega(y_s)\,ds\Bigr)\Bigm| y_t = y,\ y_u\Bigr].
\end{equation}

\begin{proposition}[It\^o factorization] \label{prop:factor}
For every model of the form \eqref{eq:gen}, $\tau = u-t$ and real frequency $\omega$, the following identity holds
\begin{equation} \label{eq:factor}
\psi(\omega,y_u ; x,y)  =  \exp \Bigl\{ \iu \omega \bigl[x  +  (r_d-r_f)\tau\bigr] + \iu\omega\rho\bigl[F(y_u)-F(y)\bigr]\Bigr\}\,H_\omega(y,y_u).
\end{equation}
It also holds for complex $\omega$ whenever $\EE[e^{-\operatorname{Im}(\omega)x_u}\mid x_t = x,y_t = y]<\infty$, with conditional quantities interpreted for almost every $y_u$.
\end{proposition}

\begin{proof}
Write $W^S = \rho W^y + \sqrt{1-\rho^2}\,W^\perp$, with $W^\perp$ a Brownian motion independent of $W^y$. Integrating \eqref{eq:gen} and using the It\^o formula for $F$,
\begin{equation*}
x_u = x + (r_d-r_f)\tau-\tfrac12\int_t^uV(y_s)\,ds + \rho\Bigl[F(y_u)-F(y)-\int_t^u\ell(y_s)\,ds\Bigr] + \sqrt{1-\rho^2}\int_t^u\sqrt{V(y_s)}\,dW^\perp_s.
\end{equation*}
Given the path of $y$, the last term is Gaussian with mean zero and variance $(1-\rho^2)\int_t^uV(y_s)\,ds$. So $\EE[e^{\iu\omega x_u}\mid(y_s)_{t\le s\le u}]$ is the exponential prefactor of \eqref{eq:factor} times $\exp(-\int_t^u\Upsilon_\omega(y_s)\,ds)$. Conditioning on $y_u$ gives \eqref{eq:factor}.
\end{proof}

The endpoint phase is explicit. The model dependence is confined to the bridge functional $H_\omega$. Table~\ref{tab:ito} lists $F$ and $\ell$ for the four models. In the Heston model $F = v/\xi$, and the phase is linear in $v_u$ as in \cite{AndersenItkinKazbek2026}. In the 3/2 model $F = \xi^{-1}\log v$, and the phase is a power of $v_u$.

\begin{table}[!htb]
\centering
\caption{It\^o coordinate $F$ and its drift $\ell$ from \eqref{eq:itocoord}. The variance is $V = v$ for the first three models and $V = \sigma^2$ for the lognormal-volatility model.}
\label{tab:ito}
\begin{tabular}{lcccc}
\toprule
Model & State $y$ & $\eta(y)$ & $F(y)$ & $\ell(y)$ \\
\midrule
Heston & $v$ & $\xi\sqrt{v}$ & $v/\xi$ & $\kappa(\theta-v)/\xi$ \\
3/2 & $v$ & $\xi v^{3/2}$ & $\xi^{-1}\log v$ & $\kappa\theta/\xi-(\kappa/\xi + \xi/2)\,v$ \\
GARCH diffusion & $v$ & $\xi v$ & $2\sqrt{v}/\xi$ & $\kappa(\theta-v)/(\xi\sqrt{v})-\xi\sqrt{v}/4$ \\
Lognormal volatility & $\sigma$ & $\varepsilon\sigma$ & $\sigma/\varepsilon$ & $0$ \\
\bottomrule
\end{tabular}
\end{table}

\subsection{The fixed-frequency backward equation} \label{sec:freqwise}

The volatility integral requires integrals of $\Psi$ against functions of $y_u$. These integrals solve one backward equation per frequency. For a complex frequency $\omega$ define the operator
\begin{equation} \label{eq:Lomega}
\mathcal{L}_\omega = \tfrac12\eta^2\partial_{yy} + \bigl(\mu + \iu\omega\rho\sqrt{V}\eta\bigr)\partial_y-\tfrac12(\omega^2 + \iu\omega)V + \iu\omega(r_d-r_f).
\end{equation}

\begin{proposition}[Fixed-frequency equation] \label{prop:freqwise}
Let $h$ be measurable on $\mathcal{Y}$ and $\omega$ a complex frequency. Suppose, for the states and positive lags under consideration, that
\begin{equation} \label{eq:weightedmoment}
\EE\bigl[e^{-\operatorname{Im}(\omega)(x_{t + \tau}-x_t)}\bigl(1 + |h(y_{t + \tau})|\bigr)\mid y_t = y\bigr]<\infty.
\end{equation}
For bounded $h$ this follows from the corresponding moment of the spot. Define
\[
G_h(\tau,y; \omega)  =  \EE\bigl[ e^{\iu\omega(x_{t + \tau}-x_t)} h(y_{t + \tau}) \mid y_t = y\bigr].
\]
Then
\begin{equation} \label{eq:Gint}
\int_{\mathcal{Y}}\Psi(\omega,y_u ; x,y)\,h(y_u)\,dy_u = e^{\iu\omega x}\,G_h(u-t,y; \omega).
\end{equation}
When these expectations are locally bounded and the usual regularity conditions for the Feynman--Kac representation hold, $G_h$ solves
\begin{equation} \label{eq:freqwise}
\partial_\tau G_h = \mathcal{L}_\omega G_h, \qquad G_h(0,y ; \omega) = h(y).
\end{equation}
Moreover $G_h = e^{\iu\omega(r_d-r_f)\tau-\iu\omega\rho F(y)}\,U$, where $U$ solves the gauge form
\begin{equation} \label{eq:gauge}
\partial_\tau U = \tfrac12\eta^2\partial_{yy}U + \mu\,\partial_yU-\Upsilon_\omega U, \qquad U(0,y) = e^{\iu\omega\rho F(y)}h(y).
\end{equation}
\end{proposition}

\begin{proof}
Condition \eqref{eq:weightedmoment} justifies the integration of \eqref{eq:Psi} against $h$ and gives \eqref{eq:Gint}. For unbounded $h$, truncate it by $h_n = h\mathbf{1}_{\{|h|\le n\}}$ and pass to the expectation by dominated convergence. The function $e^{\iu\omega x}G_h(T-t,y ; \omega)$ is a conditional expectation of a function of $(x_T,y_T)$, so it solves the backward Kolmogorov equation of \eqref{eq:gen}. Applying the generator of $(x,y)$ to $e^{\iu\omega x}G_h$ replaces $\partial_x$ by $\iu\omega$ and gives \eqref{eq:freqwise}. Substituting $G_h = e^{\iu\omega(r_d-r_f)\tau-\iu\omega\rho F}U$ and using $\eta F' = \sqrt{V}$ cancels the term $\iu\omega\rho\sqrt{V}\eta\,\partial_y$ and turns the zero-order term into $-\Upsilon_\omega$, which gives \eqref{eq:gauge}.
\end{proof}

The gauge form \eqref{eq:gauge} is the Feynman--Kac equation of the bridge functional \eqref{eq:bridgefun}, with the original real drift $\mu$. The correlation drops out of the drift and enters through the initial data and both the factors $1-\rho^2$ and $\rho\ell$ in the potential. On the fundamental strip the zero-order term of $\mathcal{L}_{\widetilde\omega}$ damps wherever $V$ is large, because $\operatorname{Re}(\widetilde\omega^2 + \iu\widetilde\omega) = \omega^2-\chi^2 + \chi>0$ for real $\omega$ and $0<\chi<1$.

\subsection{Facets in the It\^o coordinate} \label{sec:facets}

At each date $u$ we partition $\mathcal{Y}$ into segments $[y_m,y_{m + 1})$ and approximate $x^*(u,\cdot)$ by a function that is linear in a chosen coordinate on each segment. Each linear piece below is called a facet.

In \cite{AndersenItkinKazbek2026} the exercise boundary is piecewise linear in the variance. \Cref{prop:factor} suggests another coordinate. Suppose that on a segment $[y_m,y_{m + 1})$ of a partition of $\mathcal{Y}$ the boundary is linear in $F$, so that $x^*(u,y_u) = a_m + b_mF(y_u)$. Then the payoff transform \eqref{eq:Ptilde} contains the factors $e^{(\iu\omega-\chi)b_mF(y_u)}$ and $e^{(1 + \iu\omega-\chi)b_mF(y_u)}$.

\begin{corollary}[Volatility integral for It\^o-linear facets] \label{cor:facets}
Let $\tau = u-t$ and, for $k = 1,2$,
\begin{equation*}
h_k(y_u) = e^{(k-1 + \iu\omega-\chi)b_mF(y_u)}\,\mathbf{1}_{[y_m,y_{m + 1})}(y_u).
\end{equation*}
Assume the hypotheses of \cref{thm:collapse} for almost every $y_u$ in the segment, the weighted integrability condition \eqref{eq:weightedmoment} for $h_1,h_2$ at $\widetilde\omega$, and joint absolute integrability of the spectral pairing on the segment and frequency band. Then
\begin{align} \label{eq:facetint}
\int_{y_m}^{y_{m + 1}}J(u,y_u ; x,y)\,dy_u &= \mathbf{1}_{ \{\chi>0\}}\,r_d\, \mathbb{P}\bigl(y_u\in[y_m,y_{m + 1})\mid y_t = y\bigr) \\
 &+ \frac{1}{2 \pi} \int_\Omega e^{\iu \widetilde \omega x + (\iu \omega - \chi) a_m} \Bigl[ \frac{ r_d\,G_{h_1}( \tau,y; \widetilde\omega)}{ \iu \omega - \chi} - \frac{r_f\,e^{a_m}\,G_{h_2}(\tau,y;\widetilde\omega)}{1 + \iu \omega - \chi} \Bigr] d \omega. \nonumber
\end{align}
In the gauge form \eqref{eq:gauge} the initial data are $e^{\iu\widetilde\omega\rho F}h_k$, a single exponential of $F$ on the segment.
\end{corollary}

\begin{proof}
Integrate \eqref{eq:parseval} over the segment, exchange the order of integration and apply \eqref{eq:Gint}.
\end{proof}

The facet factor and the It\^o phase combine into a single exponential of $F(y_u)$. For a single facet over $\mathcal{Y}$, the volatility integral is therefore a Laplace transform in the It\^o coordinate. In the Heston model $F = v/\xi$, and this transform is exponential-affine in the initial variance, which recovers the tilted transforms and the M\"obius propagator of \cite{AndersenItkinKazbek2026}. In the 3/2 model $F = \xi^{-1}\log v$, and it is a Mellin transform in $v_u$. In the lognormal-volatility model it is a Laplace transform in $\sigma_u$, and in the GARCH diffusion a Laplace transform in $\sqrt{v_u}$. Whether the boundary is in fact closer to linear in $F$ than in $V$ is an empirical question.

\subsection{Negative moments} \label{sec:moments}

Condition (H0) is a moment condition on the spot. On the fundamental strip it always holds by \cref{thm:collapse}. For $\chi<0$ it requires the negative moment $\EE[S_u^{\chi}]$ to be finite, and non-affine models can violate this for every negative $\chi$.

\begin{proposition}[Negative moments in the GARCH and lognormal-volatility models] \label{prop:negmom}
With positive volatility of volatility, and with $\kappa,\theta>0$ for the GARCH diffusion, $\EE[S_u^{\chi}\mid x_t = x,\ y_t = y] = \infty$ for every $\chi<0$ and every $u>t$ in both the lognormal-volatility model and the GARCH diffusion.
\end{proposition}

\begin{proof}
The argument of \cref{prop:factor}, with $\iu\omega$ replaced by $\chi$, gives
\begin{equation*}
\EE\bigl[S_u^\chi\bigm|(y_s)_{t\le s\le u}\bigr] = S_t^\chi e^{\chi(r_d-r_f)\tau}\exp\Bigl\{\chi\rho\bigl[F(y_u)-F(y)\bigr]-\chi\rho\int_t^u\ell(y_s)\,ds + c_\chi\int_t^uV(y_s)\,ds\Bigr\},
\end{equation*}
with $c_\chi = \tfrac12\bigl(\chi^2(1-\rho^2)-\chi\bigr)>0$. In the lognormal-volatility model, $\ell = 0$ and $F(\sigma) = \sigma/\varepsilon$. Let $s_0 = (t + u)/2$. Given $\sigma_{s_0}$, the volatility on $[s_0,u]$ is $\sigma_{s_0}Z$, where $Z$ is a driftless geometric Brownian motion with $Z_{s_0} = 1$, independent of $\sigma_{s_0}$. The event $\mathcal{E} = \{\int_{s_0}^uZ_s^2\,ds\ge\tau/4,\ Z_u\le2\}$ has a probability $\pi_{\mathcal{E}}>0$ that does not depend on $\sigma_{s_0}$. On $\mathcal{E}$, $\int_t^u\sigma_s^2\,ds\ge\sigma_{s_0}^2\tau/4$ and $\sigma_u\le2\sigma_{s_0}$. Hence
\begin{equation*}
\EE[S_u^\chi]\ge C_0\,\pi_{\mathcal{E}}\,\EE\Bigl[\exp\bigl(c_\chi\tau\sigma_{s_0}^2/4-2|\chi\rho|\sigma_{s_0}/\varepsilon\bigr)\Bigr] = \infty,
\end{equation*}
with $C_0 = S_t^\chi e^{\chi(r_d-r_f)\tau-\chi\rho\sigma/\varepsilon}$, because $\sigma_{s_0}$ is lognormal.

For the GARCH diffusion, shift the initial time to zero, write $B_s = W^v_{t + s}-W^v_t$ and $a = \kappa + \xi^2/2$, and use the explicit solution
\begin{equation} \label{eq:garchpath}
v_{t + s} = e^{\xi B_s-as}\Bigl(v + \kappa\theta\int_0^s e^{-\xi B_z + az}\,dz\Bigr), \qquad 0\le s\le\tau.
\end{equation}
Fix $\delta>0$, let $b_L(s) = L\min(2s/\tau,1)$, and define $\mathcal E_L = \{\sup_{0\le s\le\tau}|B_s-b_L(s)|\le\delta\}$ for $L>0$. The Cameron--Martin formula, using $b_L' = 2L/\tau$ on $[0,\tau/2]$ and zero afterwards, gives
\begin{equation} \label{eq:garchtube}
\mathbb P(\mathcal E_L)\ge p_\delta\exp\bigl[-(L^2 + 2\delta L)/\tau\bigr], \qquad p_\delta = \mathbb P\bigl(\sup_{0\le s\le\tau}|B_s|\le\delta\bigr)>0.
\end{equation}
On $\mathcal E_L$, \eqref{eq:garchpath} implies $v_{t + s}\ge c>0$ throughout the interval, $v_{t + s}\le C e^{\xi L}$ throughout the interval, and $v_{t + s}\ge c e^{\xi L}$ on its second half. Here and below the positive constants do not depend on $L$. Consequently $\int_t^u v_s\,ds\ge c e^{\xi L}$, while $|F(v_u)| + \int_t^u|\ell(v_s)|\,ds\le C(1 + e^{\xi L/2})$. The latter bound controls the $v^{-1/2}$ term in $\ell$ as well as the terminal $\sqrt v$ term. The conditional-moment identity above and \eqref{eq:garchtube} therefore yield
\begin{equation} \label{eq:garchnegmoment}
\EE[S_u^\chi]\ge C_1\exp\bigl[c_1e^{\xi L}-C_2e^{\xi L/2}-(L^2 + 2\delta L)/\tau\bigr] \longrightarrow\infty \qquad (L\to\infty),
\end{equation}
which proves the claim.
\end{proof}

Heuristically, in both models a high-volatility excursion can persist over an interval of fixed length, because the diffusion coefficient of the logarithm of volatility is constant. Such paths make positive exponential moments of integrated variance diverge, as the preceding proof shows. The 3/2 model behaves differently. Its drift $-\kappa v^2$ and its volatility of variance $\xi v^{3/2}$ make the characteristic duration of an excursion to a high level $\bar v$ scale as $1/\bar v$. Negative moments of the 3/2 model can then be finite.

\section{The 3/2 model} \label{sec:32}

In this model we take $\kappa,\theta,\xi>0$ and the variance dynamics
\begin{equation} \label{eq:model32}
dv  =  \kappa v(\theta - v)\, dt  +  \xi v^{3/2}\, dW^v.
\end{equation}
The process $w = 1/v$ is a Cox--Ingersoll--Ross (CIR) process, $dw  =  (\kappa  +  \xi^2 - \kappa\theta w)\, dt - \xi\sqrt{w}\, dW^v$. Its Feller condition $2(\kappa + \xi^2)\ge\xi^2$ holds for every $\kappa\ge0$, so $v$ neither explodes nor reaches zero. Under the candidate stock-measure change, localized before exit from compact variance intervals, the drift of $v$ becomes $\kappa\theta v-(\kappa-\rho\xi)v^2$, and $1/v$ is again a CIR process, now with constant drift $\kappa-\rho\xi + \xi^2$. The discounted spot is a true martingale exactly when this process cannot reach zero, which gives $\rho\xi\le\kappa + \xi^2/2$, \cite{CarrSun2007,Drimus2012}. For large positive $\rho\xi$ relative to $\kappa$, the 3/2 price is therefore a strict local martingale. By \cref{rem:martingale} the EEP form remains valid, while the DEP form and the Put--Call duality need the condition.

\subsection{The conditional CF} \label{sec:cf32}

For the 3/2 model the conditional CF stays in closed form, with $\log v_u$ in place of $v_u$. Apply It\^o's formula to $F(v) = \xi^{-1}\log v$, the It\^o coordinate of Table~\ref{tab:ito}. This gives
\begin{equation} \label{eq:ito32}
\int_t^u \sqrt{v_s}\, dW^v_s  =  \frac{1}{\xi}\log\frac{v_u}{v_t} - \frac{\kappa\theta(u-t)}{\xi}  +  \Bigl(\frac{\kappa}{\xi}  +  \frac{\xi}{2}\Bigr)\mathcal{I}, \qquad \mathcal{I}  =  \int_t^u v_s\, ds.
\end{equation}
Conditional on the variance path, $x_u$ is Gaussian with variance $(1-\rho^2)\mathcal{I}$. Its mean depends on the path only through $\mathcal{I}$ and the endpoints. Hence
\begin{equation} \label{eq:psi32}
\psi(\omega, v_u ; x, v)  =  \Bigl(\frac{v_u}{v}\Bigr)^{\iu\omega\rho/\xi} e^{\iu\omega \bar x}\, \EE\bigl[ e^{-\hat\gamma(\omega)\mathcal{I}} \mid v_t  =  v,\ v_u \bigr].
\end{equation}
Here $\bar x  =  x  +  \int_t^u (r_d - r_f)\, ds - \rho\kappa\theta(u-t)/\xi$ collects the deterministic drift. The tilt is $\hat\gamma(\omega)  =  \tfrac12\omega^2(1-\rho^2) - \iu\omega\bigl[\rho(\kappa/\xi  +  \xi/2) - \tfrac12\bigr]$. This is \cref{prop:factor} with $\Upsilon_\omega(v) = \iu\omega\rho\kappa\theta/\xi + \hat\gamma(\omega)v$. It is the Heston structure with $v_u$ replaced by $\log v_u$.

The bridge factor is also explicit. With $w = 1/v$, the integrated variance is $\mathcal{I}  =  \int_t^u ds/w_s$. The reciprocal-Bessel transform of \cite{PitmanYor1982} gives the order shift used for the 3/2 model by \cite{Baldeaux2012ExactSimulation}, and \cite{ZhengZeng2016PartialTransform} also treat a time-dependent mean-reversion level. In the domain specified below,
\begin{equation} \label{eq:bridge32}
\EE\bigl[ e^{-\hat\gamma\mathcal{I}} \mid v_t, v_u \bigr]  =  \frac{I_{\nu(\omega)}(z)}{I_\nu(z)},
\end{equation}
where $I_\nu$ is the modified Bessel function of the first kind and
\begin{equation*}
\nu  =  1  +  \frac{2\kappa}{\xi^2}, \qquad \nu(\omega)  =  \sqrt{\nu^2  +  \frac{8\hat\gamma(\omega)}{\xi^2}}, \qquad z  =  \frac{2\kappa\theta}{\xi^2 \sinh\bigl(\kappa\theta(u-t)/2\bigr)\sqrt{v_t v_u}}.
\end{equation*}
The square root is chosen with nonnegative real part and $\nu(0) = \nu$, with boundary values obtained by continuity from the admissible domain. This is the 3/2 counterpart of the Broadie--Kaya construction, \cite{BroadieKaya2006}, with a frequency-dependent order. The next result specifies the bridge domain, including the critical tilt.

\begin{proposition}[Domain of the reciprocal-CIR bridge transform] \label{prop:bridgedomain32}
Fix $t<u$ and positive endpoints $v_t,v_u$. For real $\gamma$, the bridge expectation $B(\gamma) = \EE[e^{-\gamma\mathcal I}\mid v_t,v_u]$ is finite if and only if $\gamma\ge\gamma_c$, where $\gamma_c = -\xi^2\nu^2/8$. On this domain,
\begin{equation} \label{eq:realbridge32}
B(\gamma) = \frac{I_{\sqrt{\nu^2 + 8\gamma/\xi^2}}(z)}{I_\nu(z)}, \qquad B(\gamma_c) = \frac{I_0(z)}{I_\nu(z)}<\infty.
\end{equation}
The same representation holds for complex $\gamma$ with $\operatorname{Re}\gamma\ge\gamma_c$, using the square-root branch specified above.
\end{proposition}

\begin{proof}
For real $\gamma>\gamma_c$, the absolute-continuity relation between Bessel processes of indices $\nu$ and $\sqrt{\nu^2 + 8\gamma/\xi^2}>0$, followed by the CIR time change and conditioning on both endpoints, gives \eqref{eq:realbridge32}, see \cite{PitmanYor1982,Baldeaux2012ExactSimulation}. Monotone convergence as $\gamma\downarrow\gamma_c$ proves the finite critical value displayed there.

To prove necessity, use $\left.\partial_a I_a(z)\right|_{a = 0} = -K_0(z)<0$ for $z>0$, where $K_0$ is the modified Bessel function of the second kind, \cite{NIST2010}. The right derivative of $B$ obtained from \eqref{eq:realbridge32} tends to $-\infty$ as $\gamma\downarrow\gamma_c$. If $B(\gamma_0)$ were finite for some $\gamma_0<\gamma_c$, then $\mathcal I e^{-\gamma_c\mathcal I}\le C e^{-\gamma_0\mathcal I}$ would make that derivative finite by dominated convergence, a contradiction. For complex $\gamma$, the formula follows by analytic continuation on $\operatorname{Re}\gamma>\gamma_c$ and dominated convergence on its boundary, since $e^{-\gamma_c\mathcal I}$ is integrable under the bridge.
\end{proof}

A Monte Carlo check on a test CIR case agrees with the closed form. The check integrates \eqref{eq:bridge32} against the CIR transition law of the endpoint, which gives $\EE[e^{-\hat\gamma\mathcal{I}}]$ in closed form through a Laplace--Mellin integral of the Bessel function. For $dw = (3-2w)\,dt + \sqrt{w}\,dW$ with $w_0 = 1.5$, horizon $0.5$ and $\hat\gamma = 0.3$, the closed form gives $0.895334$. Simulation with $2\times10^5$ paths and exact CIR transitions on 400 steps gives $0.895342\pm0.000118$ (95\% band).

The log-price part of DSINC therefore applies without change. For each $v_u$, the inner integral remains a single spectral pairing of $\Psi(\widetilde\omega, v_u) = \psi(\widetilde\omega,v_u)\,p_y(v_u\mid v)$ with $\widetilde P$.

\subsection{The outer variance integral} \label{sec:outer}

For a full-range facet linear in $\log v$, the outer variance integral has a closed form through Kummer functions. In the Heston model, the integral of $\Psi$ against an exponential of $v_u$ is exponential-affine in the initial variance. A facet $x^* =  a_m  +  b_m v_u$ adds the factor $e^{(\iu\omega-\chi) b_m v_u}$, and the affine transform absorbs it.

In the 3/2 model, $\psi$ carries the power $v_u^{\iu\omega\rho/\xi}$. A facet linear in $v$ then gives a power times an exponential times the kernel, which has no closed form.

The remedy is to make the facets linear in $\log v$, so that $x^*  =  a_m  +  b_m \log v_u$. These are the facets of \cref{cor:facets}. The facet then contributes a power of $v_u$. The full-range integral becomes a Mellin transform of the order-shifted CIR kernel,
\begin{equation} \label{eq:kummer}
\int_0^\infty w^{a-1} e^{-cw} I_{\nu'}\bigl(2\delta\sqrt{w}\bigr)\, dw  =  \frac{\delta^{\nu'}\, \Gamma(a  +  \nu'/2)}{c^{a + \nu'/2}\, \Gamma(\nu' + 1)}\; {}_1F_1\Bigl(a  +  \frac{\nu'}{2};\ \nu'  +  1;\ \frac{\delta^2}{c}\Bigr),
\end{equation}
valid for $\operatorname{Re} c > 0$ and $\operatorname{Re}(a  +  \nu'/2) > 0$. Here $\Gamma$ is the gamma function and ${}_1F_1$ is the Kummer function. The identity holds for complex $a$ and complex order $\nu'$, as a numerical check confirms. The Kummer function takes the place of the M\"obius update.

\begin{proposition}[Facet transform of the 3/2 model] \label{prop:facet32}
Let $\tau = u-t>0$, $v>0$, and define
\begin{equation} \label{eq:jointdomain32}
\mathcal D(\chi) = \nu^2 + \frac{8\hat\gamma(-\iu\chi)}{\xi^2} = \Bigl(\nu-\frac{2\rho\chi}{\xi}\Bigr)^2 + \frac{4\chi(1-\chi)}{\xi^2}.
\end{equation}
For complex $\omega,\lambda$, put $\chi_\omega = -\operatorname{Im}\omega$. Assume
\begin{equation} \label{eq:facetdomain32}
\mathcal D(\chi_\omega)\ge0, \qquad \operatorname{Re}\lambda<1 + \frac{\nu + \sqrt{\mathcal D(\chi_\omega)}}{2}-\frac{\rho\chi_\omega}{\xi}.
\end{equation}
These conditions are equivalent to absolute integrability of $e^{\iu\omega(x_u-x_t)}v_u^\lambda$. Set
\begin{equation*}
C = \frac{2\kappa\theta}{\xi^2(1-e^{-\kappa\theta\tau})}, \qquad X = \frac{2\kappa\theta}{\xi^2v(e^{\kappa\theta\tau}-1)}, \qquad \alpha = \frac{\nu(\omega)-\nu}{2} + \frac{\iu\omega\rho}{\xi}.
\end{equation*}
With the square-root branch of \cref{prop:bridgedomain32} and powers of positive real quantities defined using their real logarithms,
\begin{equation} \label{eq:facet32}
\EE\bigl[e^{\iu\omega(x_u-x_t)}\,v_u^{\lambda}\bigm| v_t  =  v \bigr]  =  e^{\iu \omega(r_d-r_f) \tau}\, C^{\lambda} X^{\alpha}\, \frac{\Gamma \bigl(1  +  \nu(\omega) - \alpha - \lambda \bigr)}{\Gamma \bigl(1  +  \nu(\omega) \bigr)}\; {}_1F_1\bigl(\alpha  +  \lambda;\ 1  +  \nu(\omega);\ -X\bigr).
\end{equation}
At $\lambda = 0$ this is the CF of the 3/2 model given in \cite{Lewis2000,CarrSun2007}. Values obtained by analytic continuation outside \eqref{eq:facetdomain32} need not represent expectations.
\end{proposition}

\begin{proof}
The modulus of the random integrand is $e^{\chi_\omega(x_u-x_t)}v_u^{\operatorname{Re}\lambda}$. Its conditional expectation given the terminal variance is a positive endpoint factor times the bridge transform with real tilt $\hat\gamma(-\iu\chi_\omega)$. By \cref{prop:bridgedomain32}, that bridge transform is finite precisely when $\mathcal D(\chi_\omega)\ge0$. Write $d = \sqrt{\mathcal D(\chi_\omega)}$ when this holds. Multiplication by the CIR transition density shows that the remaining integrand, in $w_u = 1/v_u$, is asymptotic to a positive constant times $w_u^{(\nu + d)/2-\chi_\omega\rho/\xi-\operatorname{Re}\lambda}$ as $w_u\downarrow0$, and has exponential decay as $w_u\to\infty$. This proves the claimed equivalence with \eqref{eq:facetdomain32}, including $d = 0$.

For the complex transform, multiply \eqref{eq:psi32} and \eqref{eq:bridge32} by $v_u^\lambda = w_u^{-\lambda}$ and by
\begin{equation*}
p_w(w_u\mid w) = Ce^{-C(we^{-\kappa\theta\tau} + w_u)}\Bigl(\frac{w_u}{we^{-\kappa\theta\tau}}\Bigr)^{\nu/2}I_\nu\bigl(2C\sqrt{ww_ue^{-\kappa\theta\tau}}\bigr), \qquad w = 1/v.
\end{equation*}
The Bessel factor $I_\nu$ cancels. The integral is \eqref{eq:kummer} with $a = 1 + \nu/2-\iu\omega\rho/\xi-\lambda$, $\nu' = \nu(\omega)$, $\delta = C\sqrt{we^{-\kappa\theta\tau}}$ and $c = C$. The bound $\operatorname{Re}\nu(\omega)\ge d$ makes $\operatorname{Re}(a + \nu'/2)>0$ a consequence of \eqref{eq:facetdomain32}. Collecting powers and applying ${}_1F_1(a';b';X) = e^X{}_1F_1(b'-a';b';-X)$ gives \eqref{eq:facet32}. At $\lambda = 0$, writing $\bar p = -\kappa + \iu\omega\rho\xi$ gives $\alpha = -(1/2-\bar p/\xi^2) + [(1/2-\bar p/\xi^2)^2 + (\omega^2 + \iu\omega)/\xi^2]^{1/2}$ and $1 + \nu(\omega) = 2(\alpha + 1-\bar p/\xi^2)$, the parameters of the known formula.
\end{proof}

For one facet $x^* = a_m + b_m\log v_u$ over the whole variance range, \cref{cor:facets} needs \eqref{eq:facet32} at the reflected frequency, with $\lambda = (\iu\omega - \chi)b_m$ and $\lambda = (1 + \iu\omega-\chi)b_m$. The outer integral is then in closed form when both transforms satisfy \eqref{eq:facetdomain32}. Admissibility of the spot damping alone does not ensure integrability of an additional endpoint power.

\begin{corollary}[Admissible Put damping in the 3/2 model] \label{cor:adm32}
For $\chi<0$, condition (H0) holds if and only if
\begin{equation} \label{eq:adm32}
\nu^2 + \frac{8\hat\gamma(-\iu\chi)}{\xi^2}\ge0 \qquad\text{and}\qquad 1 + \frac{\nu + \nu(-\iu\chi)}{2}-\frac{\chi\rho}{\xi}>0.
\end{equation}
\end{corollary}

\begin{proof}
Set $\omega = -\iu\chi$ and $\lambda = 0$ in \eqref{eq:facetdomain32}. If its first condition fails, \cref{prop:bridgedomain32} gives an infinite bridge expectation for every positive pair of endpoints, and the unconditional negative moment is infinite. Otherwise, the positive endpoint integral converges precisely under the strict second inequality, by the small-$w_u$ asymptotic in the proof of \cref{prop:facet32}. Equality is allowed in the first condition because the critical bridge value is $I_0(z)/I_\nu(z)<\infty$. Equality in the second condition gives a logarithmically divergent endpoint integral.
\end{proof}

For example, with $\kappa = 20$, $\theta = 0.04$, $\xi = 4$ and $\rho = -0.5$, the first condition gives $\chi\ge-4.35$, and the second condition holds there. The heavy variance tail of the 3/2 model can make exponential moments of $\mathcal{I}$ infinite, so a negative $\chi$ must be checked against \eqref{eq:adm32}. No such check is needed on the fundamental strip.

\myparagraph{Segments.}
The band-mass framework carries over. The weights come from the exact 3/2 law, a noncentral chi-square law in $1/v$. The restricted transforms follow from a Mellin inversion over each segment, the analogue of the Fourier formula in the appendix of \cite{AndersenItkinKazbek2026}. Log coordinates suit the 3/2 model because the variance spans orders of magnitude. If the boundary is closer to linear in $\log v$, the band-mass bias also shrinks.

An alternative keeps the facets linear in $v$ and integrates $v_u$ by Gauss quadrature on each segment, using the closed-form tilted density. This computes the segment integrals directly, so it has no band-mass bias. Its cost lies in Bessel functions of complex order, which remain affordable if $I_{\nu(\omega)}(z)$ is tabulated in $z$ per frequency and lag.

\myparagraph{Time-dependent coefficients.}

With constant $\kappa$, $\xi$ and $\rho$, a deterministic $\theta(t)$ preserves the reciprocal-CIR construction, \cite{ZhengZeng2016PartialTransform}. Define $\Theta(t,u) = \int_t^u\kappa\theta(s)\,ds$ and $\mathcal{T}(t,u) = \int_t^u e^{\Theta(t,s)}\,ds$. The formulas hold with
\begin{equation*}
C = \frac{2e^{\Theta(t,u)}}{\xi^2\mathcal{T}(t,u)}, \qquad X = \frac{2}{\xi^2v\,\mathcal{T}(t,u)}, \qquad z = 2\sqrt{CX/v_u},
\end{equation*}
while the deterministic leverage phase uses $\rho\,\Theta(t,u)/\xi$ in place of $\rho\kappa\theta\tau/\xi$. The Bessel order and the moment conditions are unchanged.

Piecewise-constant $\xi$ and $\rho$, as in \cite{AndersenItkinKazbek2026}, break closure at the breakpoints. Let $\Delta_k$ be the jump of $\rho/\xi$ at a breakpoint $t_k$. That breakpoint contributes the factor $v_{t_k}^{-\iu\widetilde\omega\Delta_k}$, a power where the Heston model has an exponential shear. The Bessel order also changes between intervals. The Chapman--Kolmogorov equation restores the computation with one one-dimensional quadrature in $v_{t_k}$ per breakpoint crossed. With three intervals there are at most two crossings. \Cref{prop:freqwise} offers a direct alternative. On each interval the fixed-frequency equation has constant coefficients, and a breakpoint only switches the operator.

\myparagraph{Cost and oscillation.}
Special functions with complex order or parameters cost far more than exponentials and hyperbolic functions. In the Heston model the transform depends on the conditioning variance $v$ only through the factor $e^{\calA  +  \calB v}$, where $\calA$ and $\calB$ are the coefficients of the tilted CIR transform. Additional variance nodes therefore add little cost. The closed-form evaluation in the 3/2 model requires one evaluation per node. Tabulation can reduce repeated evaluations. A solver of the boundary equation with Kummer facets has not been implemented here. Whether the spectral integrand is non-oscillatory depends on the bridge law of $\mathcal{I}$ near zero. It also depends on the extra phase from the power of $v_u$.

\section{The SABR model} \label{sec:sabr}

The generalized integral transform of \cite{ItkinMuravey2021SABR} allows the early-exercise formulation of \cite{AndersenItkinKazbek2026} to be extended to SABR dynamics, which have no affine CF. We use that construction to obtain the transition operator, and we keep the exercise surface in the integration limits of the premium. A modification of the DSINC pairing accommodates absorption at zero. Throughout this section the transition operator is not restricted by the exercise decision, while absorption at the model boundary remains in force.

\subsection{Model and transition representation} \label{sec:sabrmodel}

Following \cite{ItkinMuravey2021SABR}, consider a forward price $f$ with
\begin{equation} \label{eq:sabr}
df_t = \sigma_tf_t^{\beta}\,dW^f_t, \qquad d\sigma_t = -\kappa(t)\sigma_t\,dt + \varepsilon(t)\sigma_t\,dW^\sigma_t, \qquad d\langle W^f,W^\sigma\rangle_t = \rho(t)\,dt.
\end{equation}
Here $\beta\in(0,1)$ is the elasticity of \cite{Hagan2002}, $\kappa(t)$ is a drift rate of the volatility, $\varepsilon(t)$ is the volatility of volatility and $\rho(t)$ is the correlation. The coefficients are deterministic, sufficiently regular and bounded on the pricing horizon, with $\varepsilon>0$ and $|\rho|<1$. For $\kappa = 0$ and constant coefficients this is the classical SABR model. We impose absorption at $f = 0$. The volatility remains positive at finite times when started from $\sigma>0$. The nonnegative local martingale $f$ has a finite first moment, even when equality of that moment to its initial value has not been established.

We price a cash-settled American Put on the forward, with payoff $(K-f)^ + $ and a deterministic discount rate $r$. In the notation of \cref{sec:theory} this is the case $r_d = r_f = r$, with log-moneyness $x = \log(f/K)$ and payoff weight $g(x) = r(1-e^x)$. An exchange rate with a rate differential also fits \eqref{eq:sabr}. Its forward has a volatility with a linear drift, and its exercise payoff then has a time-dependent strike.

Set $\nu_\beta = 1/(2-2\beta)$ and $R = f^{1-\beta}/(1-\beta)$. In the variables $(R,\sigma)$ the undiscounted backward generator becomes
\begin{equation} \label{eq:sabrgen}
\mathcal{L}_t = \sigma^2\Bigl(\frac12\partial_{RR}-\frac{2\nu_\beta-1}{2R}\,\partial_R\Bigr) + \frac12\varepsilon(t)^2\sigma^2\partial_{\sigma\sigma}-\kappa(t)\sigma\,\partial_\sigma + \rho(t)\varepsilon(t)\sigma^2\partial_R\partial_\sigma.
\end{equation}
For $\rho = 0$, with $J_\nu$ the Bessel function of the first kind, the eigenfunctions
\begin{equation} \label{eq:besseleig}
\Phi_q(R) = R^{\nu_\beta}J_{\nu_\beta}(qR), \qquad q\ge0,
\end{equation}
diagonalize the first term, with eigenvalue $-q^2/2$. We call $q$ the Bessel frequency. The realization that vanishes at $R = 0$ represents the positive-price part of the absorbed process. Let $\mathcal{A}_{t,u} = \int_t^u\sigma_s^2\,ds$ be the integrated variance. For a bounded function $h$ of the volatility, define the tilted volatility transform
\begin{equation} \label{eq:Ghat}
\widehat G_h(t,\sigma ; u,q) = \EE\bigl[e^{-q^2\mathcal{A}_{t,u}/2}\,h(\sigma_u)\bigm|\sigma_t = \sigma\bigr].
\end{equation}
For each $q\ge0$ it solves a one-dimensional backward equation,
\begin{equation} \label{eq:Ghatpde}
\partial_t\widehat G_h + \frac12\varepsilon(t)^2\sigma^2\partial_{\sigma\sigma}\widehat G_h-\kappa(t)\sigma\,\partial_\sigma\widehat G_h-\frac12q^2\sigma^2\widehat G_h = 0, \qquad \widehat G_h(u,\sigma ; u,q) = h(\sigma).
\end{equation}
With $h$ a Dirac mass at $\sigma_u$, $\widehat G_h$ becomes the tilted kernel, a density in $\sigma_u$ that we write $\widehat G_{\sigma_u}(t,\sigma ; u,q)$. The continuous joint transition density of $(R_u,\sigma_u)$ with respect to $dR_u\,d\sigma_u$ is then
\begin{equation} \label{eq:sabrdensity}
p^ + (R_u,\sigma_u\mid R,\sigma) = R_u^{\,1-2\nu_\beta}\int_0^\infty q\,\Phi_q(R)\,\Phi_q(R_u)\,\widehat G_{\sigma_u}(t,\sigma ; u,q)\,dq.
\end{equation}
This is the half-line spectral representation with weight $R_u^{1-2\nu_\beta}$ and spectral measure $q\,dq$. It follows by conditioning on the volatility path, because the Bessel process runs on the clock $\mathcal{A}$. The formula gives only the positive-price density. Its missing probability is the mass at $f_u = 0$. In a numerical implementation, the spectral integral and the volatility domain are truncated separately. Equivalently, the volatility propagator can be applied to terminal functions, without constructing a delta-function kernel explicitly. For a function $H(R_u,\sigma_u)$ with the required integrability this reads
\begin{equation} \label{eq:sabrkernel}
\EE\bigl[H(R_u,\sigma_u)\mathbf{1}_{\{R_u>0\}}\bigm|R_t = R,\ \sigma_t = \sigma\bigr] = \int_0^\infty q\,\Phi_q(R)\,\widehat G_{\widehat H(q,\cdot)}(t,\sigma ; u,q)\,dq,
\end{equation}
where $\widehat H(q,\sigma_u) = \int_0^\infty R_u^{\,1-2\nu_\beta}\,\Phi_q(R_u)\,H(R_u,\sigma_u)\,dR_u$ is the Hankel transform with weight $R_u^{1-2\nu_\beta}$.

Let $x_u = \log(f_u/K)$. On $\{f_u>0\}$ the joint density of $(x_u,\sigma_u)$ is
\begin{equation} \label{eq:sabrlogdensity}
p(x_u,\sigma_u\mid x,\sigma) = (Ke^{x_u})^{1-\beta}\,p^ + \bigl(R(Ke^{x_u}),\sigma_u\bigm| R(Ke^x),\sigma\bigr),
\end{equation}
because $dR/dx = f^{1-\beta}$. This is the density $p$ of \cref{sec:theory} with $y = \sigma$, restricted to positive prices. Thus the Bessel representation supplies the transform $\Psi$ that DSINC requires. It does not supply an affine dependence on the initial price, and conversion to the Fourier representation has a computational cost. In particular, an interchange of the Bessel and Mellin integrals is not assumed without a separate convergence argument.

\subsection{A DSINC pairing that includes the mass at zero} \label{sec:sabrpairing}

For the Put the exercise region is $0\le f\le f^*(u,\sigma_u)$, with $f^*(u,\sigma_u) = Ke^{x^*(u,\sigma_u)}>0$. The integrand weight is $G(f_u, \sigma_u)  =  r(1-f_u/K) \mathbf{1}_{\{f_u\le f^*(u,\sigma_u)\}}$, and the premium integrand $J$ of \eqref{eq:J} now includes the absorbed mass. Negative damping of the complete log-price law is unavailable when absorption has positive probability, because the corresponding negative moment is infinite. Instead, subtract $G(0,\sigma_u) = r$. On $\{f_u>0\}$ the residual is
\begin{equation} \label{eq:sabrresidual}
G(Ke^{x_u},\sigma_u)-r = -r\,e^{x_u}\mathbf{1}_{\{x_u\le x^*(u,\sigma_u)\}}-r\,\mathbf{1}_{\{x_u>x^*(u,\sigma_u)\}}.
\end{equation}
This residual tends to zero at the absorbing boundary. Its transform against $e^{(\iu\omega-\chi)x_u}$ exists for $0<\chi<1$ and equals \eqref{eq:Ptilde} with $r_d = r_f = r$. The algebraic expression is the same as in the unregularized Put pairing, but its contour and the added constant differ. The transform $\Psi(\widetilde\omega,\sigma_u ; x,\sigma)$ of \eqref{eq:sabrlogdensity} is the expectation of $(f_u/K)^{\chi-\iu\omega}\mathbf{1}_{\{f_u>0,\ \sigma_u\in d\sigma_u\}}$ per unit $d\sigma_u$. No normalization by the volatility density is needed. The required positive fractional moment is finite because $s^\chi\le1 + s$ for $s\ge0$ and $0<\chi<1$.

\begin{proposition}[Regularized DSINC pairing] \label{prop:sabrpairing}
Fix $t<u$, $f,\sigma>0$ and $0<\chi<1$. Suppose that, for almost every $\sigma_u$, the damped density $p_\chi(x_u,\sigma_u) = e^{\chi x_u}p(x_u,\sigma_u\mid x,\sigma)$ belongs to $L^1(\mathbb{R})\cap L^2(\mathbb{R})$ in $x_u$, and that the spectral pairing below is absolutely integrable jointly in $(\omega,\sigma_u)$. Then
\begin{equation} \label{eq:sabrfull}
\int_0^\infty J(u,\sigma_u ; x,\sigma)\,d\sigma_u = r + \frac{1}{2\pi}\int_0^\infty\int_{\mathbb{R}}\Psi(\widetilde\omega,\sigma_u ; x,\sigma)\,\widetilde P(\omega,\sigma_u)\,d\omega\,d\sigma_u.
\end{equation}
For the band $\Omega$, let $p_{\chi,\Omega}$ be the Fourier projection of $p_\chi$ onto $\Omega$, and set $c_j(\sigma_u) = e^{-\chi x_j}p_{\chi,\Omega}(x_j,\sigma_u)$. Its damped-sinc representation is
\begin{equation} \label{eq:sabrsinc}
e^{-\chi x_u}p_{\chi,\Omega}(x_u,\sigma_u) = \sum_{j\in\mathbb{Z}}c_j(\sigma_u)\,\phi_j(x_u),
\end{equation}
with the basis $\phi_j$ of \cref{sec:collapse}. The coefficients collapse, in the $L^2(\Omega)$ sense, according to
\begin{equation} \label{eq:sabrcollapse}
\sum_{j\in\mathbb{Z}}c_j(\sigma_u)\,e^{-(\iu\omega-\chi)x_j} = \frac{1}{\Delta x}\,\Psi(\widetilde\omega,\sigma_u ; x,\sigma), \qquad \omega\in\Omega.
\end{equation}
Consequently, pairing the projected density with the residual \eqref{eq:sabrresidual} gives the finite-band version of \eqref{eq:sabrfull}, with $\mathbb{R}$ replaced by $\Omega$.
\end{proposition}

\begin{proof}
At $f = 0$ the residual $G-r$ vanishes, so the volatility integral of $J$ equals $r$ plus the integral of the residual against the positive-price density. The function $e^{-\chi x_u}$ times the residual belongs to $L^1\cap L^2$ for $0<\chi<1$. Integrating its two pieces gives \eqref{eq:Ptilde}. Fourier pairing with $p_\chi$ and the assumed integrability give \eqref{eq:sabrfull}. Shannon reconstruction of $p_{\chi,\Omega}$ gives \eqref{eq:sabrsinc}, and its Fourier-series identity on $\Omega$ gives \eqref{eq:sabrcollapse}. Pairing in the damped variables then yields the finite-band formula.
\end{proof}

\Cref{prop:sabrpairing} specializes \cref{thm:collapse}, \cref{rem:bandlimit} and \cref{rem:residue} to an absorbed price. It applies to any transition law that satisfies its assumptions, including correlated SABR. It does not assert that the actual density is band-limited. If $\Psi$ is replaced by a numerical approximation $\widehat\Psi$, the resulting finite-band pairing $\widehat{\mathcal{J}}_\Omega$ satisfies
\begin{equation} \label{eq:sabrerror}
\Bigl|\int_0^\infty J\,d\sigma_u-\widehat{\mathcal{J}}_\Omega\Bigr|\le E_{\mathrm{tail}} + \frac{1}{2\pi}\int_0^\infty\int_\Omega\bigl|\Psi-\widehat\Psi\bigr|\,\bigl|\widetilde P\bigr|\,d\omega\,d\sigma_u,
\end{equation}
where $E_{\mathrm{tail}} = (2\pi)^{-1}\int_0^\infty\int_{\mathbb{R}\setminus\Omega}|\Psi\widetilde P|\,d\omega\,d\sigma_u$ is \eqref{eq:tail} integrated over the terminal volatility. This estimate separates frequency truncation from transform approximation for a prescribed exercise surface. It does not include quadrature, surface interpolation or errors in solving the nonlinear boundary equation.

\subsection{The premium integrand in the Hankel representation} \label{sec:sabrcarry}

Inserting \eqref{eq:sabrlogdensity} into \eqref{eq:sabrfull} requires the Fourier transform of $\Phi_q$ in the log-price, a Mellin transform that is an analytically continued Weber integral, \cite{NIST2010}. For a fixed Bessel frequency the pairing over $\omega$ need not converge absolutely, because $\widetilde P$ decays only like $1/\omega$. The $\omega$-integral can instead be carried out exactly. By Parseval's identity it returns the residual \eqref{eq:sabrresidual} in the price variable, whose Hankel transform has a closed form.

\begin{proposition}[Premium integrand in the Hankel representation]  \label{prop:hankelcarry}
Let $\rho = 0$, let $R = R(Ke^x)$, and let $R^*(\sigma_u) = R\bigl(f^*(u,\sigma_u)\bigr)$ be the boundary in the Bessel coordinate. Then
\begin{equation} \label{eq:sabrJ}
\int_0^\infty J(u,\sigma_u ; x,\sigma)\,d\sigma_u = r + \int_0^\infty q\,\Phi_q(R)\,\widehat G_{\widehat P(q,\cdot)}(t,\sigma ; u,q)\,dq,
\end{equation}
where
\begin{equation} \label{eq:sabrPhat}
\widehat P(q,\sigma_u) = - \frac{r}{q} \Bigl[ \frac{(1-\beta)^{2\nu_\beta}}{K}\, R^*(\sigma_u)^{1 + \nu_\beta}J_{\nu_\beta + 1}\bigl(qR^*(\sigma_u)\bigr) + R^*(\sigma_u)^{1-\nu_\beta}J_{\nu_\beta-1}\bigl(qR^*(\sigma_u)\bigr)\Bigr].
\end{equation}
The European Put follows with $r$ replaced by $1$ and $R^*$ by $R(K)$, and with the result multiplied by $Ke^{-r(T-t)}$.
\end{proposition}

\begin{proof}
By the first step in the proof of \cref{prop:sabrpairing}, the volatility integral of $J$ equals $r$ plus the integral of the residual against the positive-price density. In the Bessel coordinate $f = ((1-\beta)R)^{2\nu_\beta}$, and the residual equals $-rf/K$ for $R\le R^*$ and $-r$ for $R>R^*$. Apply \eqref{eq:sabrkernel}. The identity $\frac{d}{dz}\bigl[z^{\nu + 1}J_{\nu + 1}(z)\bigr] = z^{\nu + 1}J_\nu(z)$ gives $\int_0^{R^*}R_u^{1 + \nu_\beta}J_{\nu_\beta}(qR_u)\,dR_u = R^{*\,1 + \nu_\beta}J_{\nu_\beta + 1}(qR^*)/q$. The identity $\frac{d}{dz}\bigl[z^{1-\nu}J_{\nu-1}(z)\bigr] = -z^{1-\nu}J_\nu(z)$, together with the decay of $z^{1-\nu}J_{\nu-1}(z)$ at infinity for $\nu>1/2$, gives the improper integral
\[
\int_{R^*}^\infty R_u^{1-\nu_\beta}J_{\nu_\beta}(qR_u)\,dR_u = R^{*\,1-\nu_\beta}J_{\nu_\beta-1}(qR^*)/q .
\]
Here $\nu_\beta>1/2$ for every $\beta\in(0,1)$. Combining the two pieces gives \eqref{eq:sabrPhat}.
\end{proof}

The initial forward enters \eqref{eq:sabrJ} only through $\Phi_q(R)$, so it can be varied cheaply during the computation of the boundary. The propagation is in the volatility alone, one equation per Bessel frequency, which is the Bessel analogue of \cref{prop:freqwise}. Near $q = 0$ the integrand of \eqref{eq:sabrJ} behaves like $q^{2\nu_\beta-1}$. It is smooth for $\beta = 1/2$, where $\nu_\beta = 1$, and otherwise the quadrature must be graded toward zero. At a time lag $\tau = u-t$ the Bessel frequencies must reach several multiples of $1/(\sigma\sqrt{\tau})$, and the integrand oscillates with a period near $2\pi/(R + R^*)$. Short lags therefore need many Bessel frequencies.

\subsection{The American exercise equation} \label{sec:sabramer}

Assume a single exercise boundary $f^*(t,\sigma)$ and the regularity and integrability conditions for the early-exercise-premium representation of \cite{AndersenItkinKazbek2026}. Since the generator annihilates $K-f$ in the exercise region, the normalized cost of carry is $g(x)  =  r(1-e^x)$, the case $r_d = r_f = r$ of \eqref{eq:J}. The price and the boundary then satisfy
\begin{align} \label{eq:sabrprice}
P(t,f,\sigma) &= P_E(t,f,\sigma) + K\int_t^Te^{-\int_t^ur(s)\,ds}\int_0^\infty J\bigl(u,\sigma_u ; \log(f/K),\sigma\bigr)\,d\sigma_u\,du, \\
K-f^*(t,\sigma) &= P_E\bigl(t,f^*(t,\sigma),\sigma\bigr) + K\int_t^Te^{-\int_t^ur(s)\,ds}\int_0^\infty J\bigl(u,\sigma_u ; \log(f^*(t,\sigma)/K),\sigma\bigr)\,d\sigma_u\,du. \nonumber
\end{align}
The European expectation uses the same regularized pairing, with $r$ replaced by $1$ and $x^*$ by $0$, multiplied by $Ke^{-\int_t^Tr(s)ds}$. Thus absorption is included in both the European term and the premium. For other contracts or a different pricing drift, the cost of carry must be recomputed from $(r - \partial_u - \mathcal{L}_u) h$, where $h$ is the exercise payoff. In particular, the cost of carry of a spot exchange rate must not be transferred unchanged to a driftless-forward specification.

The surface is computed backward in time on a volatility grid. During its nonlinear solution the transition operator is reused, while the exercise level changes inside $\widetilde P$ or $\widehat P$. The prescribed barrier of \cite{ItkinMuravey2021SABR} is not replaced by $f^*(t,\sigma)$ inside the moving-domain transform, because a volatility-dependent integration endpoint would produce additional derivative terms. The fixed-domain transition construction avoids those terms.

For constant $r$, let $t_n = n\Delta t$ for $n = 0,\dots,N$, with $\Delta t = T/N$, and use the trapezoidal rule in $u$, with weights $\Delta t_k$ equal to $\Delta t$ for $k<N$ and $\Delta t/2$ for $k = N$. We discretize \eqref{eq:Ghatpde} in $\zeta = \log\sigma$ on nodes $\sigma_i$ by second-order differences with Neumann ends. This gives a real matrix $\mathbf{A}_q$ for each Bessel frequency and the one-step propagator $\mathbf{M}_q = \exp(\Delta t\,\mathbf{A}_q)$. Let $\widehat P_k(q)$ be the vector of \eqref{eq:sabrPhat} at the nodes, built from the boundary at $t_k$, and let $\widehat P^E(q)$ be its European counterpart. The premium accumulates in
\begin{equation} \label{eq:sabrQ}
Q_n(q) = \sum_{k = n + 1}^{N}\Delta t_k\,e^{-r(t_k-t_n)}\,\mathbf{M}_q^{\,k-n}\,\widehat P_k(q) = e^{-r\Delta t}\,\mathbf{M}_q\bigl(\Delta t_{n + 1}\widehat P_{n + 1}(q) + Q_{n + 1}(q)\bigr),
\end{equation}
with $Q_N = 0$. Let $q_j$ and $w_j$ be quadrature nodes and weights in the Bessel frequency, let $\mathbf{e}$ be the vector of ones, and let $D_n = r\sum_{k = n + 1}^N\Delta t_k\,e^{-r(t_k-t_n)}$. Value matching at the node $\sigma_i$ then reads, with $R = R(f)$,
\begin{align} \label{eq:sabrvm}
1-\frac{f}{K} &= e^{-r(T-t_n)}\Bigl\{1 + \sum_jw_jq_j\Phi_{q_j}(R)\bigl[\mathbf{M}_{q_j}^{\,N-n}\mathbf{e}\bigr]_i\,\widehat P^E(q_j)\Bigr\} \\
 &+ \tfrac14\Delta t\,r\Bigl(1-\frac{f}{K}\Bigr) + D_n + \sum_jw_jq_j\Phi_{q_j}(R)\bigl[Q_n(q_j)\bigr]_i, \qquad f = f^*(t_n,\sigma_i). \nonumber
\end{align}
The term $\Delta t\,r(1-f/K)/4$ uses the right-hand limit of the premium integrand as $u\downarrow t_n$, which is $g/2$ at a regular, nondegenerate exercise boundary. At short times a diffusion started on a regular boundary lies below it with probability one half. This does not imply a smooth time integrand. For the time-homogeneous tests, we solve \eqref{eq:sabrvm} by bisection in $R$, simultaneously at all nodes, with a narrow bracket\footnote{The bisection starts from a bracket, an interval in which the residual changes sign, chosen narrow and just below the boundary of the next date.} just below the boundary of the next date. The exercise region shrinks as the remaining maturity increases, so this rule follows the boundary branch adjoining the continuation region. With time-dependent coefficients both bracket directions must be allowed.

\subsection{Correlation and computational scope} \label{sec:sabrcorr}

For $\rho\ne0$, the mixed derivative in \eqref{eq:sabrgen} couples the Bessel modes. One option is the perturbative construction of \cite{ItkinMuravey2021SABR}. For time-dependent correlation, write $\rho(t)=c\,\bar\rho(t)$ with a constant $c$, and expand an undiscounted transition problem as $w=\sum_{n\ge0}c^nw_n$. If $\mathcal{L}_{0,t}$ is \eqref{eq:sabrgen} with $\rho=0$, then
\begin{equation} \label{eq:sabrpert}
(\partial_t+\mathcal{L}_{0,t})w_0=0, \qquad (\partial_t+\mathcal{L}_{0,t})w_n=-\bar\rho(t)\varepsilon(t)\sigma^2\,\partial_R\partial_\sigma w_{n-1}, \qquad n\ge1 .
\end{equation}
The zeroth term carries the prescribed terminal data, and the higher terms have zero terminal data. Expanding in the constant $c$ avoids unaccounted time derivatives of powers of $\rho(t)$. Each correction uses the uncorrelated propagator with a known source, subject to the regularity needed for the expansion. Accuracy at the target correlation requires a truncation study, and convergence at $c=1$ is not assumed.

\begin{myremark}[Correlation destroys the Bessel-frequency decoupling] \label{rem:coupling}
The eigenfunctions \eqref{eq:besseleig} satisfy $\partial_R\Phi_q(R) = qR^{\nu_\beta}J_{\nu_\beta-1}(qR)$, by the identity $J_\nu'(z) = J_{\nu-1}(z)-\nu J_\nu(z)/z$. This function has Bessel order $\nu_\beta-1$ and is not a multiple of any single $\Phi_{q'}$. Its Hankel transform is spread over all Bessel frequencies, so the mixed term $\rho\varepsilon\sigma^2\partial_{R\sigma}$ couples them. Independent one-dimensional volatility equations per frequency are therefore an exact feature of the uncorrelated construction $\rho=0$, and not of arbitrary correlated SABR.
\end{myremark}

Consequently the Bessel representation of \cref{sec:sabrmodel} does not extend to $\rho \ne 0$ without either a mode-coupling treatment or an approximation. \Cref{app:corr} writes the expansion \eqref{eq:sabrpert} in integral form as a Duhamel iteration, and a heat-kernel approximation is an alternative for short maturities, see \cref{app:sabrheat}. Neither has been assessed numerically here.

A change of volatility coordinate avoids the coupling. Let $x = \log(f/K)$ and $y = \sigma f^{\beta-1}$, the instantaneous lognormal volatility of the forward. The scaling $f\mapsto\lambda f$, $\sigma\mapsto\lambda^{1-\beta}\sigma$ with $\lambda>0$ leaves \eqref{eq:sabr} invariant, shifts $x$ by $\log\lambda$ and leaves $y$ unchanged. This reduction is due to \cite{Park2013}, who prices European options with the variables $X = \log f$ and $Z = f^{1-\beta}/\sigma = 1/y$ by uncoupled one-dimensional equations after a Fourier decomposition in $X$. Our contribution is its use inside the regularized early-exercise representation and the boundary recursion. It\^o's formula gives
\begin{equation} \label{eq:sabrxy}
dx_t = -\tfrac12 y_t^2\,dt + y_t\,dW^f_t, \qquad d\log y_t = \bigl(\tfrac12(1-\beta)y_t^2-\kappa(t)-\tfrac12\varepsilon(t)^2\bigr)dt + \varepsilon(t)\,dW^\sigma_t-(1-\beta)\,y_t\,dW^f_t,
\end{equation}
so the coefficients do not depend on $x$. The pair $(x,y)$ is a model of the form \eqref{eq:gen} with $V(y) = y^2$, with two differences. The correlation of $x$ and $y$ depends on $y$, and $y$ explodes at the absorption time $\tau_0 = \inf\{s\ge t: f_s = 0\}$, whereas \eqref{eq:gen} assumes a non-explosive volatility factor. The proof of \cref{prop:freqwise} uses only the independence of the coefficients from $x$, so \eqref{eq:freqwise} holds with $\rho\sqrt{V}\eta$ replaced by $d\langle x,y\rangle_t/dt = \rho(t)\varepsilon(t)y^2-(1-\beta)y^3$, provided the expectation is taken for the process killed at $\tau_0$,
\[
G_h(\tau,y; \omega) = \EE\bigl[e^{\iu\omega(x_{t + \tau}-x_t)}h(y_{t + \tau})\,\mathbf{1}_{\{\tau_0>t + \tau\}}\mid y_t = y\bigr].
\]
This is the positive-price part of the regularized pairing of \cref{prop:sabrpairing}, written in $y_u$ instead of $\sigma_u$, a change of variables at fixed $x_u$. Absorbed paths enter only through the residue term, because the damped factor $e^{\chi(x_u-x_t)}$ vanishes as $f_u\to0$ on the fundamental strip. The killed solution vanishes as $y\to\infty$. At the upper end of a truncated grid we keep a homogeneous Neumann condition, as for the Bessel frequencies in \cref{sec:sabramer}, and our tests show that neither this choice nor the position of that end affects the prices. Correlated SABR therefore needs one volatility equation per Fourier frequency. Its fixed-frequency solution is an alternative to a two-dimensional FD solution, of comparable accuracy in our tests.

The Bessel construction and the fixed-frequency solution for model \eqref{eq:gen} share one property that concerns accuracy rather than speed. The initial price enters only analytically, through $\Phi_q(R)$ for SABR and through $e^{(\chi-\iu\omega)x}$ for model \eqref{eq:gen}. Prices at all spot levels therefore follow from the same accumulated vectors without interpolation, and the exercise boundary is a continuous root rather than a quantity resolved on a price grid. For SABR in the coordinates \eqref{eq:sabrxy}, a change of $f$ at fixed $\sigma$ also moves $y$ and needs interpolation in $y$. A two-dimensional FD solver also gives prices at all grid points, but off-grid values require interpolation and the boundary is resolved only to the mesh width. The correlated iteration of \cref{app:corr} retains this property, because each of its orders applies the uncorrelated propagator. The property may matter for Greeks and for boundaries that must be differentiated. We do not quantify it here.

\subsection{Numerical tests} \label{sec:sabrnum}

We take $\beta = 1/2$, so $\nu_\beta = 1$, with $\sigma_0 = 0.3$, $\varepsilon = 0.3$, $\kappa = 0$, $r = 0.05$, $K = 1$ and $T = 0.5$. The Bessel frequencies use composite Gauss--Legendre rules, with 8 nodes on each interval of length $1/2$ up to $q = 150$ for European prices. For American prices they use 4 nodes on each such interval, up to $q = 600$ with $N = 50$ and up to $q = 850$ with $N = 100$.

First, with constant volatility the model is the absorbed CEV model, whose density is known in closed form. Then $\widehat G$ reduces to $e^{-q^2\sigma^2\tau/2}$. For $\beta = 1/2$, $\sigma = 0.3$ and $\tau = 0.5$, \cref{prop:hankelcarry} reproduces direct quadrature of the closed-form density to eight digits, both at $f = K = 1$ and at $f = K = 0.05$, where the absorption probability is $0.108$. For $\beta = 0.3$ the ungraded rule is off by $8.5\times10^{-5}$, and a geometric grading toward $q = 0$ restores eight digits.

Second, European Puts with stochastic volatility. Given the volatility path, the forward follows the absorbed CEV law on the clock $\mathcal{A}$, so conditional Monte Carlo needs only the closed-form CEV price at the simulated clock. With $2\times10^5$ paths and 1000 exact volatility steps, it gives $0.138680(39)$, $0.082703(44)$ and $0.044861(41)$ for $f/K = 0.9$, $1.0$ and $1.1$, respectively, with the half-widths of the 95\% Monte Carlo confidence intervals reported in parentheses\footnote{The parenthetical number is expressed in units of the last displayed decimal place, here $10^{-6}$. Thus, $0.138680(39)$ denotes the estimate $0.138680$ with confidence half-width $39\times10^{-6} = 0.000039$, corresponding to the interval $[0.138641,\,0.138719]$. The other two half-widths are $0.000044$ and $0.000041$. Under the normal approximation, the 95\% confidence half-width equals $1.96$ times the estimated standard error of the Monte Carlo mean. These intervals quantify sampling uncertainty; they do not account for time-discretization bias or other systematic numerical errors.}. \Cref{prop:hankelcarry} gives $0.138699$, $0.082724$ and $0.044881$. At $f = K = 0.05$ the two give normalized prices $P_E/K = 0.358771(179)$ and $0.358856$, and the absorption probabilities $0.11476$ and $0.11489$. The volatility grid here spans $\log\sigma_0\pm5\varepsilon\sqrt{T}$ with 121 nodes.

Third, American Puts, in Table~\ref{tab:sabr}. We call the evaluation of \cref{prop:hankelcarry} with the recursion \eqref{eq:sabrQ} the Hankel method. The reference solves the problem in $(f,\zeta)$ on $f\in[0,3]$, with value one at $f = 0$, where the Put is exercised at absorption. It uses four implicit Euler half-steps, Crank--Nicolson steps and the operator splitting of \cite{IkonenToivanen2004}. This is a standard second-order scheme. All derivatives, the mixed one included, are replaced by second-order central differences, as in \cite{inTHoutFoulon2010}, here on a uniform grid, and the value at $f = 3$ is zero. The implicit Euler half-steps of \cite{Rannacher1984} damp the oscillations that Crank--Nicolson steps, \cite{CrankNicolson1947}, produce at the kink of the payoff. Each step solves the full two-dimensional system with one sparse LU factorization, computed once and without dimensional splitting, and the prices at the spot levels follow by cubic interpolation in $f$. This scheme is chosen for the testing purposes only. For more sophisticated FD schemes suitable for production use, see \cite{Andersen2025} and the references therein.

Both methods use the volatility range $\log\sigma_0\pm3.5\varepsilon\sqrt{T}$ with Neumann ends, with 81 nodes for the Hankel method and the fine grid and 41 for the coarse grid. Refining the grid raises the FD prices by $1.5\times10^{-5}$ to $1.9\times10^{-5}$. The Hankel method with $N = 100$ lies within $8\times10^{-6}$ of the fine prices, on the side toward which refinement moves them, and within $2\times10^{-6}$ of the second-order extrapolation. Doubling $N$ changes its prices by $10^{-6}$. The Hankel method is slower in this implementation. Elapsed times in this paper are medians of three runs on one core of a virtualized Intel Xeon processor, with Python 3.12, NumPy 2.4 and SciPy 1.17. The Hankel method takes 30~s for $N = 50$ and 81~s for $N = 100$, mostly for Bessel-function evaluations in the computation of the boundary with 4800 and 6800 frequencies, against 1.5~s for the fine grid.

\begin{table}[htbp]
\centering
\caption{American Puts on the SABR forward with $\beta = 1/2$ and $\rho = 0$, prices in units of the strike. FD denotes the FD reference in $(f,\zeta)$, on $301\times41$ nodes with 150 steps (coarse) and $601\times81$ nodes with 300 steps (fine). The Hankel method uses \cref{prop:hankelcarry} with $N$ time steps. The last column is its early-exercise premium for $N = 100$.}
\label{tab:sabr}
\begin{tabular}{cccccc}
\toprule
$f/K$ & FD coarse & FD fine & Hankel, $N = 50$ & Hankel, $N = 100$ & EEP \\
\midrule
0.9 & 0.139639 & 0.139655 & 0.139663 & 0.139662 & 0.000963 \\
1.0 & 0.083101 & 0.083120 & 0.083129 & 0.083128 & 0.000404 \\
1.1 & 0.045018 & 0.045033 & 0.045040 & 0.045039 & 0.000158 \\
\bottomrule
\end{tabular}
\end{table}

The tests above concern the Hankel implementation at $\rho = 0$. Numerical results for correlated SABR, obtained by solving the fixed-frequency equation in the coordinates \eqref{eq:sabrxy}, are reported in \cref{sec:general}, with comparisons against FD in \cref{tab:extra}.

\section{General models: fixed-frequency numerical solution} \label{sec:general}

For a model of the form \eqref{eq:gen}, the fixed-frequency equation evaluates the required terminal-volatility integral directly when a practical closed-form transform is unavailable. The collapse still holds in that case, but the speed advantage over FD is no longer automatic. This section describes the numerical solution, the representation of the boundary and the time recursion, and then reports numerical tests.

We solve \eqref{eq:freqwise} on a uniform grid $\zeta_1<\dots<\zeta_{N_y}$ in the log-volatility $\zeta = \log y$. In this coordinate the diffusion coefficient of the volatility factor is $\eta_\zeta = \eta/y$, its drift is $\mu_\zeta = \mu/y-\eta^2/(2y^2)$, and \eqref{eq:freqwise} becomes
\begin{equation} \label{eq:freqzeta}
\partial_\tau G = \tfrac12\eta_\zeta^2\,\partial_{\zeta\zeta}G + \bigl(\mu_\zeta + \iu\omega\rho\sqrt{V}\eta_\zeta\bigr)\partial_\zeta G-\tfrac12(\omega^2 + \iu\omega)\,V\,G + \iu\omega(r_d-r_f)\,G.
\end{equation}
Second-order central differences and homogeneous Neumann conditions at the ends of the grid give a complex tridiagonal matrix $\mathbf{A}_\omega$. The ends lie far in the tails of the volatility distribution. On the fundamental strip the zero-order term damps at high volatility, so the Neumann conditions have little effect. The coefficients do not depend on time, so the one-step propagator $\mathbf{M}_\omega = \exp(\Delta t\,\mathbf{A}_\omega)$ is computed once per frequency by scaling and squaring, \cite{MolerVanLoan2003,Higham2005}. Here $\Delta t$ is the time step defined below. Piecewise-constant coefficients need one exponential per interval.

\subsection{Two models without a closed form} \label{sec:noclosed}

The GARCH diffusion and the lognormal-volatility model have no closed-form volatility integral, and they are the examples of this section. For the GARCH diffusion, $\zeta = \log v$, $\eta_\zeta = \xi$ and $\mu_\zeta = \kappa\theta e^{-\zeta}-\kappa-\xi^2/2$, so \eqref{eq:freqzeta} reads
\begin{equation} \label{eq:freqgarch}
\partial_\tau G = \tfrac12\xi^2\partial_{\zeta\zeta}G + \bigl(\kappa\theta e^{-\zeta}-\kappa-\tfrac12\xi^2 + \iu\omega\rho\xi e^{\zeta/2}\bigr)\partial_\zeta G-\tfrac12(\omega^2 + \iu\omega)e^{\zeta}G + \iu\omega(r_d-r_f)G.
\end{equation}
Its gauge form \eqref{eq:gauge} carries the potential \eqref{eq:potential} with the drift $\ell$ of Table~\ref{tab:ito}, which contains $v^{-1/2}$, $v^{1/2}$ and $v$. No transformation known to us reduces it to a hypergeometric equation, and \cite{BaroneAdesiEtAl2005} price European options in this model through an approximation. For the lognormal-volatility model, $\zeta = \log\sigma$, $\eta_\zeta = \varepsilon$ and $\mu_\zeta = -\varepsilon^2/2$, so
\begin{equation} \label{eq:freqlnv}
\partial_\tau G = \tfrac12\varepsilon^2\partial_{\zeta\zeta}G + \bigl(-\tfrac12\varepsilon^2 + \iu\omega\rho\varepsilon e^{\zeta}\bigr)\partial_\zeta G-\tfrac12(\omega^2 + \iu\omega)e^{2\zeta}G + \iu\omega(r_d-r_f)G.
\end{equation}
In the gauge form the potential is $\tfrac12[\omega^2(1-\rho^2) + \iu\omega]e^{2\zeta}$, an exponential potential. The bridge functional then involves the law of the integrated variance of a geometric Brownian motion given its endpoint, which is of Hartman--Watson type, \cite{Yor1992}, and has no closed form practical for repeated evaluation. For both models, prices are therefore computed by solving \eqref{eq:freqgarch} and \eqref{eq:freqlnv} numerically for each frequency node, with the boundary sampled at the volatility nodes. The 3/2 model is computed in the same way and checked against \cref{prop:facet32}.

\subsection{Nodal boundary and time recursion} \label{sec:nodal}

Facets are unnecessary once the fixed-frequency problem is solved numerically. We represent the boundary at each date by its values $x^*(u,y_i)$ at the nodes $y_i = e^{\zeta_i}$. Assume \eqref{eq:weightedmoment} for $h = \widetilde P(\omega,\cdot)$ at the reflected frequency and joint absolute integrability of the spectral pairing in terminal volatility and frequency. Then \eqref{eq:Gint} gives
\begin{equation} \label{eq:nodalouter}
\int_{\mathcal{Y}}J(u,y_u ; x,y)\,dy_u = \mathbf{1}_{\{\chi>0\}}\,r_d + \frac{1}{2\pi}\int_\Omega e^{\iu\widetilde\omega x}\,G_{\widetilde P(\omega,\cdot)}(u-t,y ; \widetilde\omega)\,d\omega.
\end{equation}
The discrete version applies powers of $\mathbf{M}_{\widetilde\omega}$ to the vector of nodal payoff transforms\footnote{\emph{Nodal} refers to the representation of the exercise boundary by its point values at the volatility grid nodes $y_i = e^{\zeta_i}$, as opposed to a piecewise-linear (facet) parametrization. The payoff transform at each date is assembled from these nodal values.}. The boundary may then take any shape in $y$. The band-mass approximation of \cite{AndersenItkinKazbek2026}, which combines the contributions of individual facets through band-mass weights, is not needed. The remaining approximations are frequency truncation, frequency quadrature, volatility-domain truncation, spatial discretization and time quadrature.

Let $t_n = n\Delta t$ for $n = 0,\dots,N$, with $\Delta t = T/N$. We use the trapezoidal rule in $u$, with weights $\Delta t_k$ equal to $\Delta t$ for $k<N$ and to $\Delta t/2$ for $k = N$. Let $\widetilde P_k(\omega)$ be the vector of payoff transforms built from the nodal boundary $x^*(t_k,y_i)$. The spectral part of the premium accumulates in
\begin{equation} \label{eq:Qrec}
Q_n(\omega) = \sum_{k = n + 1}^{N}\Delta t_k\,e^{-r_d(t_k-t_n)}\,\mathbf{M}_{\widetilde\omega}^{\,k-n}\,\widetilde P_k(\omega),
\end{equation}
which satisfies $Q_N = 0$ and $Q_n = e^{-r_d\Delta t}\,\mathbf{M}_{\widetilde\omega}\bigl(\Delta t_{n + 1}\widetilde P_{n + 1} + Q_{n + 1}\bigr)$. Each step costs one matrix--vector product per frequency. The density $p$ is real, so the integrand at $-\omega$ is the complex conjugate of the integrand at $\omega$, and only $\omega\ge0$ is needed. Let $\omega_j$ and $\varpi_j$ be Gauss--Legendre nodes and weights on $[0,\pi/\Delta x]$, and let $D_n = r_d\sum_{k = n + 1}^{N}\Delta t_k\,e^{-r_d(t_k-t_n)}$ be the discounted residue term. The value-matching condition at the node $y_i$ then reads
\begin{equation} \label{eq:vm}
1-e^{x} = \frac{P_E(t_n,x,y_i)}{K} + \tfrac14\Delta t\,g(x) + \mathbf{1}_{\{\chi>0\}}D_n + \frac{1}{\pi}\operatorname{Re}\sum_j\varpi_j\,e^{(\chi-\iu\omega_j)x}\bigl[Q_n(\omega_j)\bigr]_i, \qquad x = x^*(t_n,y_i).
\end{equation}
The term $\Delta t\,g(x)/4$ uses the right-hand limit at the lower endpoint of the time integral. To see this, suppose $x^*$ is locally $C^{1,2}$ in time and volatility near $(t_n,y)$ and write $b_y = \partial_yx^*(t_n,y)$. The instantaneous variance of $x_u-x^*(u,y_u)$ at the starting boundary is
\begin{equation} \label{eq:normalvariance}
a_\perp = V(y)-2\rho\sqrt{V(y)}\eta(y)b_y + \eta(y)^2b_y^2 = V(y)(1-\rho^2) + \bigl(\eta(y)b_y-\rho\sqrt{V(y)}\bigr)^2>0.
\end{equation}
The short-time normal limit therefore gives
\begin{equation} \label{eq:eeprightlimit}
\lim_{u\downarrow t_n}\int_{\mathcal Y}J(u,y_u ; x^*(t_n,y),y)\,dy_u = \tfrac12 g(x^*(t_n,y)).
\end{equation}
The indicator at exactly $u = t_n$ does not determine this limit. A term of order $\sqrt{u-t_n}$ can remain in the integrand, so the endpoint correction alone does not establish second-order convergence of the time quadrature. With $\mathbf{e}$ the vector of ones, the European price is
\begin{equation} \label{eq:euronodal}
\frac{P_E(t_n,x,y_i)}{K} = e^{-r_d(T-t_n)}\Bigl\{\mathbf{1}_{\{\chi>0\}} + \frac{1}{\pi}\operatorname{Re}\sum_j\varpi_j\,\frac{e^{(\chi-\iu\omega_j)x}\bigl[\mathbf{M}_{\widetilde\omega_j}^{\,N-n}\mathbf{e}\bigr]_i}{(\iu\omega_j-\chi)(1 + \iu\omega_j-\chi)}\Bigr\}.
\end{equation}
At the initial state, with the spot in the continuation region, the price is the right-hand side of \eqref{eq:vm} without the endpoint term.

\subsection{Root solving and cost} \label{sec:root}

The exercise boundary is the upper edge of the contact region, $x^*(t,y) = \sup\{x\le0:P(t,x,y) = K(1-e^x)\}$. Value matching also holds inside that region, so it does not by itself distinguish an arbitrary interior contact point from the boundary. In \eqref{eq:vm}, the residual on the exercise side is only of order $\Delta t$. A bracket extending deep into that region can therefore lose its sign when quadrature errors exceed the residual. For $\chi<0$, the factor $e^{\chi x}$ amplifies these errors as $x$ decreases.

The time-homogeneous tests use a local bracket of fixed initial width below $x^*(t_{n + 1},y_i)$, widened where necessary to recover a sign change. This follows the boundary branch adjoining the continuation region from the next date, rather than selecting an interior contact point. Monotonicity here follows from the inclusion of admissible stopping times as the remaining maturity increases: at a fixed state, the exercise region can only shrink. Bisection is applied simultaneously at all nodes, using 34 evaluations of \eqref{eq:vm}. This local continuation rule is a numerical boundary selection, not a uniqueness theorem for the discretized equation. For time-dependent coefficients, the previous boundary remains a predictor, but temporal monotonicity cannot in general be used to constrain the bracket. Both bracket directions must be allowed.

In the present implementation, the setup forms $N_\omega$ dense matrix exponentials of size $N_y$, at a cost of $O(N_\omega N_y^3)$. The recursion costs $O(NN_\omega N_y^2)$, and each evaluation of \eqref{eq:vm} at all volatility nodes costs $O(N_\omega N_y)$. At a fixed initial volatility state $y$, the log-moneyness enters only through $e^{(\chi - \iu\omega)x}$, so no grid in $x$ is needed, and the same accumulated vectors give the price at every spot level.

For larger volatility grids, the action $e^{\Delta t\mathbf{A}_\omega}\mathbf{v}$ on a vector $\mathbf{v}$ can instead be approximated by a Krylov method, without forming or storing the dense propagator, see \cite{ItkinDF2026}. Since $\mathbf{A}_\omega$ is tridiagonal, one Arnoldi projection onto a Krylov subspace of dimension $m$ costs $O(m^2N_y + m^3)$, including orthogonalization and evaluation of the reduced matrix exponential. With one such projection per time step and frequency, the recursion costs $O(NN_\omega(m^2N_y + m^3))$. The required dimension $m$, and any additional projections, depend on the tolerance, frequency and stiffness of the operator. This provides a possible acceleration for larger grids. The timings reported below use the explicitly formed propagators.

\subsection{Numerical tests} \label{sec:numsetup}

All tests use $T = 0.5$, $r_d = 0.05$, $r_f = 0.02$ and $K = 1$, so prices are in units of the strike, with spot levels $S/K\in\{0.9,1.0,1.1\}$. The 3/2 model has $\kappa = 20$, $\theta = 0.04$, $\xi = 4$, $\rho = -0.5$ and $v_0 = 0.04$. The lognormal-volatility model has $\varepsilon = 0.8$, $\rho = -0.3$ and $\sigma_0 = 0.15$. The GARCH diffusion has $\kappa = 2$, $\theta = 0.0225$, $\xi = 1.5$, $\rho = -0.3$ and $v_0 = 0.0225$. In all three models the discounted spot is a true martingale, in the lognormal-volatility model because $\rho\le0$, \cite{Jourdain2004}. The grids are uniform in $\zeta$ on $[\log(2\times10^{-4}),\log 2]$ for the 3/2 model, $[\log\sigma_0-4,\log\sigma_0 + 3.5]$ for the lognormal-volatility model and $[\log v_0-5,\log v_0 + 4.5]$ for the GARCH diffusion. The frequency quadrature uses 12 Gauss--Legendre nodes on each panel between the breakpoints $0,2,5,10,20,40,80,160,320,640$, which gives $N_\omega = 108$ and corresponds to $\Delta x = \pi/640$. Unless stated otherwise, $N_y = 160$, $N = 100$ and $\chi = 1/2$. All computations run in Python with NumPy and SciPy on one core, and elapsed times are measured as in \cref{sec:sabrnum}.

\subsubsection{Transforms and European prices} \label{sec:euro}

Table~\ref{tab:euro} summarizes three tests: convergence of the fixed-frequency transform, agreement of European Put prices with independent references, and sensitivity to the damping exponent.

\begin{table}[!htb]
\centering
\caption{Transform and European-price tests. Prices are in units of the strike. In panel~(b), spectral prices use \eqref{eq:euronodal} with $\chi = 1/2$, and the 3/2 closed-form reference uses \eqref{eq:facet32} with $\lambda = 0$. MC denotes conditional Monte Carlo. The 2000-step estimates use $4\times10^5$ paths; a value $a(b)$ denotes the estimate $a$ with 95\% confidence half-width $b\times10^{-6}$. Panel~(c) reports price variation over the listed damping exponents.}
\label{tab:euro}
\small
\setlength{\tabcolsep}{4pt}
\begin{tabular}{lllll}
\toprule
\multicolumn{5}{l}{\textit{(a) 3/2 transform: convergence against the closed form}} \\
\midrule
$N_y$ & $60$ & $120$ & $240$ & $480$ \\
Maximum error & $1.4\times10^{-4}$ & $3.5\times10^{-5}$ & $8.7\times10^{-6}$ & $2.2\times10^{-6}$ \\
\midrule
\multicolumn{5}{l}{\textit{(b) European Put prices}} \\
\midrule
Model & Method & $S/K = 0.9$ & $S/K = 1.0$ & $S/K = 1.1$ \\
\midrule
3/2 & Spectral & 0.102557 & 0.047224 & 0.019105 \\
& Closed form & 0.102557 & 0.047224 & 0.019105 \\
\addlinespace
Lognormal vol. & Spectral & 0.094682 & 0.035822 & 0.012168 \\
& MC, 2000 steps & 0.094598(99) & 0.035748(88) & 0.012126(58) \\
& MC, 500 steps & 0.094757 & 0.035882 & 0.012192 \\
\addlinespace
GARCH diffusion & Spectral & 0.093499 & 0.033759 & 0.009631 \\
& MC, 2000 steps & 0.093425(85) & 0.033700(68) & 0.009603(39) \\
& MC, 500 steps & 0.093570 & 0.033810 & 0.009651 \\
\midrule
\multicolumn{5}{l}{\textit{(c) Sensitivity to admissible damping}} \\
\midrule
Model & \multicolumn{2}{l}{Damping exponents $\chi$} & \multicolumn{2}{l}{Observed agreement} \\
\midrule
3/2 & \multicolumn{2}{l}{$-1,\, -1/2,\, 1/4,\, 1/2,\, 3/4$} & \multicolumn{2}{l}{Same six decimal places} \\
Lognormal vol. & \multicolumn{2}{l}{$1/4,\, 1/2,\, 3/4$} & \multicolumn{2}{l}{Variation $\le 2\times10^{-6}$} \\
GARCH diffusion & \multicolumn{2}{l}{$1/4,\, 1/2,\, 3/4$} & \multicolumn{2}{l}{Variation $\le 2\times10^{-6}$} \\
\bottomrule
\end{tabular}
\end{table}

For the transform test, we compute $G_h$ with $h(v) = v^\lambda$ for $\tau = 0.5$ and $v = 0.04$, with $\chi \in \{-1/2, 1/2\}$, $\omega \in \{0, 2, 10, 30\}$ and $\lambda \in \{0, 0.3\}$, on $v \in [10^{-4}, 4]$. Panel~(a) reports the largest absolute error against \cref{prop:facet32} over these 16 cases. Doubling the number of volatility nodes reduces the error by approximately a factor of four, consistent with second-order convergence.

Panel~(b) compares European Put prices. In the 3/2 model, the spectral and closed-form prices agree to all six displayed decimal places. For the lognormal-volatility model and the GARCH diffusion, the reference is a conditional Monte Carlo estimator that simulates the volatility path and integrates the log-price analytically given that path, \cite{Willard1997}. The main reference uses 2000 log-Euler steps. For the lognormal-volatility model, direct integration of $d\sigma_s = \varepsilon\sigma_s\,dW^\sigma_s$ gives the exact identity $\int_t^u \sigma_s\,dW^\sigma_s = (\sigma_u - \sigma_t)/\varepsilon$. The integrated variance $\int_t^u \sigma_s^2\,ds$ is still approximated on the time grid. All six spectral prices lie within the corresponding 95\% Monte Carlo confidence intervals. The independent 500-step estimates are also reported; in each case, the two Monte Carlo estimates lie on opposite sides of the spectral price.

Panel~(c) checks invariance with respect to admissible damping, as required by \cref{thm:collapse}. For the 3/2 model, we use 16 frequency nodes per panel up to $\omega = 80$ and 160 volatility nodes on $v \in [10^{-4}, 4]$. The prices agree to six decimal places for damping exponents on both sides of the pole at $\chi = 0$. With the standard quadrature, the lognormal-volatility and GARCH prices vary by at most $2\times10^{-6}$ over the tested positive damping exponents.

\subsubsection{Moment explosions on truncated grids} \label{sec:mom}

\begin{figure}[!htb]
\centering
\includegraphics[width = \textwidth]{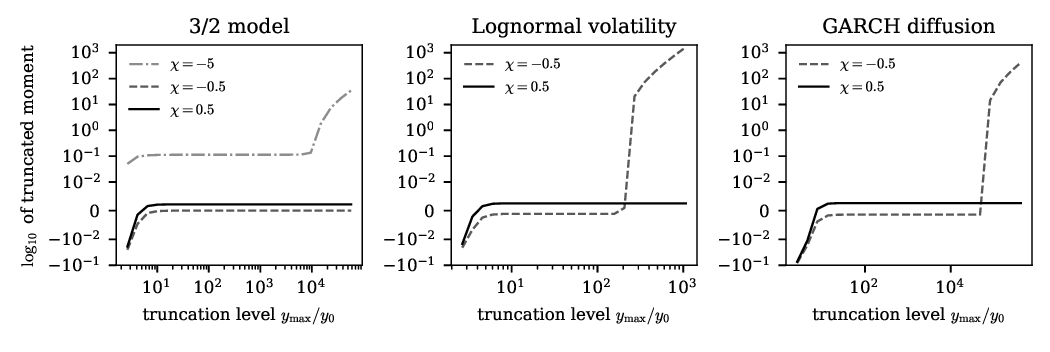}
\caption{Truncated moments $\EE[(S_T/S_0)^\chi]$ for $T = 0.5$, computed by the fixed-frequency solver with absorption at the truncation level $y_{\max}$, on a symmetric logarithmic scale. Parameters as in \cref{sec:numsetup}.}
\label{fig:moments}
\end{figure}

Figure~\ref{fig:moments} illustrates \cref{prop:negmom} and \cref{cor:adm32}. It shows the moment $\EE[(S_T/S_0)^\chi]$, computed by the fixed-frequency solver at $\widetilde\omega = -\iu\chi$ with absorption at a truncation level $y_{\max}$, as $y_{\max}$ grows. For $\chi = 1/2$ the truncated moments converge in all three models. In the 3/2 model they also converge for $\chi = -1/2$, which \cref{cor:adm32} admits. For $\chi = -5$, below the bound $-4.35$, they explode once $y_{\max}$ exceeds about $10^4v_0$. In the lognormal-volatility model the truncated moment for $\chi = -1/2$ stays at $0.99744$ up to $y_{\max}\approx270\,\sigma_0$ and then grows without bound. In the GARCH diffusion it stays at $0.99678$ up to $y_{\max}\approx8\times10^4v_0$ and then explodes. Hence a truncated grid does not reveal the failure of the Put damping at moderate truncation. The computed number is a moment of the truncated model. It depends on the truncation, and it overflows as the grid is widened. On the fundamental strip the truncated moments converge to finite limits as the truncation level increases.

\subsubsection{American Puts} \label{sec:amer}

The reference solves the two-dimensional problem in $(x,\zeta)$ on $x\in[-1.2,1.2]$. It starts with four implicit Euler half-steps, \cite{Rannacher1984}, continues with Crank--Nicolson steps, and treats the exercise constraint by the operator splitting of \cite{IkonenToivanen2004}. It is the scheme of \cref{sec:sabrnum}, with the payoff as value at $x = -1.2$, value zero at $x = 1.2$ and Neumann conditions at the ends of the $\zeta$ grid. The coarse grid has $241\times101$ nodes and 150 steps, and the fine grid $481\times201$ nodes and 300 steps. Table~\ref{tab:amer} reports the results.

Table~\ref{tab:amerchecks} summarizes the accuracy checks and computational costs. Refinement raises the FD prices, and the DSINC prices lie on the side toward which refinement moves them. If the leading FD error is of second order in the mesh, its extrapolated value is $P_{\mathrm{ext}} = P_{\mathrm{fine}} + (P_{\mathrm{fine}} - P_{\mathrm{coarse}})/3$. At comparable cost, DSINC lies closer to this extrapolation than the fine-grid FD prices. We do not claim a speed advantage beyond this setting.

\begin{table}[!htb]
\centering
\caption{American Puts, prices in units of the strike. FD denotes the finite-difference reference. DSINC uses $\chi = 1/2$, $N_y = 160$, $N_\omega = 108$ and $N$ time steps. The last column is the early-exercise premium of DSINC with $N = 100$.}
\label{tab:amer}
\small
\begin{tabular}{llccccc}
\toprule
Model & $S/K$ & FD coarse & FD fine & DSINC, $N = 100$ & DSINC, $N = 200$ & EEP \\
\midrule
3/2 & 0.9 & 0.107575 & 0.107593 & 0.107601 & 0.107599 & 0.005044 \\
    & 1.0 & 0.048618 & 0.048648 & 0.048662 & 0.048660 & 0.001438 \\
    & 1.1 & 0.019469 & 0.019490 & 0.019499 & 0.019497 & 0.000394 \\
\midrule
Lognormal & 0.9 & 0.101856 & 0.101864 & 0.101868 & 0.101867 & 0.007186 \\
     & 1.0 & 0.037145 & 0.037191 & 0.037209 & 0.037207 & 0.001387 \\
     & 1.1 & 0.012377 & 0.012397 & 0.012405 & 0.012403 & 0.000237 \\
\midrule
GARCH & 0.9 & 0.101095 & 0.101106 & 0.101111 & 0.101110 & 0.007612 \\
      & 1.0 & 0.035163 & 0.035206 & 0.035223 & 0.035221 & 0.001464 \\
      & 1.1 & 0.009850 & 0.009869 & 0.009877 & 0.009876 & 0.000246 \\
\bottomrule
\end{tabular}
\end{table}

\begin{table}[!htb]
\centering
\caption{Accuracy checks and elapsed times for the nine American Put cases in Table~\ref{tab:amer}. Price changes and differences are absolute and expressed in units of the strike. Unless stated otherwise, DSINC uses $N = 100$, $N_y = 160$ and $\chi = 1/2$. The extrapolated FD values assume second-order spatial convergence. Time ranges are across the three models.}
\label{tab:amerchecks}
\small
\setlength{\tabcolsep}{4pt}
\begin{tabular}{@{}p{0.65\linewidth}p{0.30\linewidth}@{}}
\toprule
\multicolumn{2}{l}{\textit{(a) Accuracy and parameter refinement}} \\
\midrule
Comparison & Reported difference or change \\
\midrule
FD coarse to fine & $8\times10^{-6}$ to $4.6\times10^{-5}$ \\
DSINC versus fine FD & At most $1.8\times10^{-5}$ \\
DSINC versus extrapolated FD & At most $4\times10^{-6}$ \\
DSINC time steps: $N = 100$ to $N = 200$ & At most $2\times10^{-6}$ \\
DSINC volatility nodes: $N_y = 160$ to $N_y = 240$ & At most $10^{-6}$ \\
DSINC damping: $\chi \in \{1/4, 1/2, 3/4\}$ & At most $2\times10^{-6}$ \\
3/2 damping: $\chi = -1/2$ versus $\chi = 1/2$ & Agreement to six digits \\
\midrule
\multicolumn{2}{l}{\textit{(b) Computational cost}} \\
\midrule
Calculation & Elapsed time (s) \\
\midrule
DSINC: complete run & 4.5--8.3 \\
DSINC: 108 matrix exponentials, included in the total & 2.1--4.0 \\
FD: fine grid & 7.8--8.7 \\
FD: coarse grid & 0.4--0.5 \\
\bottomrule
\end{tabular}
\end{table}

Table~\ref{tab:amerrobust} reports the bracket and time-refinement tests. The exercise-boundary bracket must remain local even on the fundamental strip. Time refinement also requires a sufficiently wide frequency band: with the band fixed, finer time steps can expose frequency-truncation error. When the band is enlarged, the 3/2 and GARCH prices show further convergence, while the lognormal-volatility changes are too small and irregular to identify a reliable order.

\begin{table}[!htb]
\centering
\caption{Bracket sensitivity and time refinement for American Puts. Bracket tests use $\chi = 1/2$ and $N = 100$. Price errors and changes are expressed in units of the strike. The frequency cutoff is the largest resolved positive frequency.}
\label{tab:amerrobust}
\small
\setlength{\tabcolsep}{4pt}
\begin{tabular}{@{}p{0.44\linewidth}p{0.51\linewidth}@{}}
\toprule
\multicolumn{2}{l}{\textit{(a) Initial exercise-boundary bracket}} \\
\midrule
Comparison & Result \\
\midrule
Widths $0.1$ and $0.3$ & Same prices to seven digits in all three models \\
Width $1.0$ & Errors up to $6\times10^{-3}$ \\
Whole log-moneyness range & Errors up to $5\times10^{-5}$ \\
\midrule
\multicolumn{2}{l}{\textit{(b) Time steps and frequency band}} \\
\midrule
Comparison & Result \\
\midrule
$N = 50, 100, 200$ & Successive price changes decrease by factors of $2.1$--$3.6$; observed orders $1.1$--$1.8$ \\
Further refinement to $N = 400$, cutoff $640$ & Lognormal-volatility price changes up to $3\times10^{-5}$ \\
Further refinement, cutoff $2560$ & Observed orders $1.0$--$1.4$ for 3/2 and GARCH; lognormal-volatility changes at most $4\times10^{-6}$, with no resolved order \\
\bottomrule
\end{tabular}
\end{table}

\begin{table}[!htb]
\centering
\caption{Heston American Put benchmark with strike $K = 10$. Prices are in currency units. DSINCn uses $N = 100$ and $N_y = 120$; both $\chi = -1/2$ and $\chi = 1/2$ give the displayed values. The FD calculation uses $401\times161$ nodes and 200 time steps.}
\label{tab:amerheston}
\small
\setlength{\tabcolsep}{5pt}
\begin{tabular}{lccccc}
\toprule
Method & $S = 8$ & $S = 9$ & $S = 10$ & $S = 11$ & $S = 12$ \\
\midrule
Reference, \cite{IkonenToivanen2009} & 2.0000 & 1.1076 & 0.5200 & 0.2137 & 0.0820 \\
DSINCn & 2.0000 & 1.1077 & 0.5202 & 0.2138 & 0.0821 \\
FD & 2.0000 & 1.1073 & 0.5196 & 0.2134 & 0.0819 \\
\bottomrule
\end{tabular}
\end{table}

\begin{figure}[!htb]
\centering
\includegraphics[width = \textwidth]{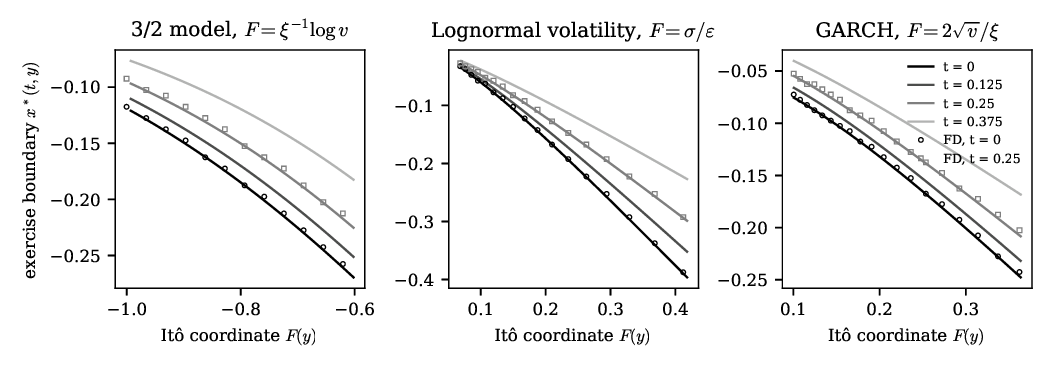}
\caption{Exercise boundaries $x^*(t,y)$ from DSINC with a nodal boundary at $t = 0$, $0.125$, $0.25$ and $0.375$ (lines, dark to light) against the It\^o coordinate $F(y)$, with the fine finite-difference boundaries at $t = 0$ (circles) and $t = 0.25$ (squares). The horizontal ranges cover the 1\% to 99\% quantiles of $y$ at $t = 0.25$.}
\label{fig:boundary}
\end{figure}

As a check of the fixed-frequency solver and the FD reference, we price the Heston benchmark of \cite{ClarkeParrott1999}, with strike $K = 10$ and the reference values reported in \cite{IkonenToivanen2009}. In this test, DSINC solves \eqref{eq:freqzeta} numerically (so we write DSINCn) for the Heston coefficients, using $N_y = 120$ variance nodes and $N = 100$ time steps, with the exercise boundary represented at the variance nodes. The closed-form tilted-CIR transform and the M\"obius propagator of \cite{AndersenItkinKazbek2026} are not used. This comparison therefore tests the numerical evaluation of the volatility integral and the boundary recursion in a model with established benchmark prices.

DSINCn gives the displayed prices for both $\chi = -1/2$ and $\chi = 1/2$. Our FD code interpolates the time value, defined as the option value minus the payoff, cubically to the spot levels and floors it at zero. Thus, the option at $S = 8$, which lies in the exercise region, receives its intrinsic value. At $S = 10$, a grid node, the FD price lies $4\times10^{-4}$ below the reference value. Using a bracket over the whole log-moneyness range gave a systematic error of about $6\times10^{-4}$ of the strike for $\chi = -1/2$, illustrating the failure described in \cref{sec:root}.

Figure~\ref{fig:boundary} shows the DSINC exercise boundaries at four dates against the It\^o coordinate $F(y)$, together with the fine FD boundaries at $t = 0$ and $t = 0.25$. On the nodes between the 2.5\% and 97.5\% quantiles of $y$ at $t = 0.25$, the two agree to within $0.0055$ in $x$, approximately one FD mesh width of $0.005$.

The boundaries are close to linear in $F$, as \cref{cor:facets} suggests. Table~\ref{tab:amerboundaryfit} compares the largest residuals from least-squares linear fits in $F(y)$ and in the variance $V(y)$ on the same nodes. The It\^o coordinate reduces these residuals by factors of 1.4 to 4.5, most strongly for the lognormal-volatility model, where the variance is the square of the state. Across the fitted nodes, the boundary spans $0.11$ to $0.29$ in $x$. The remaining curvature motivates the nodal representation of the boundary.

\begin{table}[!htb]
\centering
\caption{Largest absolute residuals in log-moneyness from least-squares linear fits to the exercise boundary. Each pair of rows compares the It\^o coordinate $F(y)$ with the variance $V(y)$ at the same date. Fits use nodes between the 2.5\% and 97.5\% quantiles of $y$ at $t = 0.25$.}
\label{tab:amerboundaryfit}
\small
\begin{tabular}{clccc}
\toprule
Date $t$ & Fitting coordinate & 3/2 & Lognormal vol. & GARCH \\
\midrule
0 & $F(y)$ & 0.0044 & 0.0063 & 0.0041 \\
  & $V(y)$ & 0.0088 & 0.0272 & 0.0080 \\
\addlinespace
0.25 & $F(y)$ & 0.0047 & 0.0045 & 0.0030 \\
     & $V(y)$ & 0.0067 & 0.0204 & 0.0078 \\
\bottomrule
\end{tabular}
\end{table}

\subsection{Further tests} \label{sec:further}

\myparagraph{Closed-form facets in the 3/2 model.}
The American prices above use the fixed-frequency solver for all three models. To test the closed-form route of \cref{sec:32}, we fit at each date one facet $x^* = a_k + b_k\log v$ to the nodal DSINC boundary of the 3/2 model, by least squares over the 1\% to 99\% quantiles of the variance at expiry. For this prescribed boundary we evaluate the premium in two ways, nodally and by \cref{cor:facets} with the facet transform \eqref{eq:facet32}, with the same 108 frequencies and time grid. The two premiums differ by at most $4\times10^{-7}$, which confirms the closed-form volatility integral in an American computation. The single facet itself is not accurate enough. It lowers the American prices by $1.8\times10^{-4}$, $4.8\times10^{-5}$ and $9.6\times10^{-6}$ for $S/K = 0.9$, $1.0$ and $1.1$, although the boundary is close to linear in $\log v$, as measured in \cref{sec:amer}. Accuracy comparable to the nodal boundary would need several facets per date, with the segment integrals of \cref{sec:outer}. With the mpmath library, the closed-form premium at one state takes 11~s, against 6~s for the nodal evaluation of the same boundary at every volatility node and spot level.

\myparagraph{Further parameter cases.}
The first eight rows of Table~\ref{tab:extra} compare DSINC with the FD reference in further cases. Three take the parameters of \cref{sec:numsetup} with maturity $T = 1$. Three take $r_d = 0.03$ and $r_f = 0.05$, so that the boundary tends to $\log(r_d/r_f)<0$ as expiry approaches, although at expiry the Put is exercised wherever its payoff is positive. One is a 3/2 model with $\kappa = 0.2$, $\xi = 1$, $\rho = 0.9$ and $\theta = v_0 = 0.04$. There $\rho\xi>\kappa + \xi^2/2$, so the discounted spot is a strict local martingale and only the EEP form applies. The last is a GARCH diffusion with $\xi = 1$ and $\rho = -0.6$ for $t<0.25$ and the parameters of \cref{sec:numsetup} afterwards. Both methods switch operators at $t = 0.25$, and the boundary bracket may move in both directions. DSINC uses $N = 200T$ steps. The FD grids have mesh widths $0.01$ and $0.005$ in $x$, $101$ and $201$ volatility nodes, and $300$ and $600$ steps per unit time, on $x\in[-1.2,1.2]$, or on $[-2.5,1.2]$ when $r_d<r_f$.

In all eight cases DSINC lies within $2.1\times10^{-5}$ of the fine grid and within $1.1\times10^{-5}$ of the second-order extrapolation. Outside the cases with $r_d<r_f$ it lies within $4\times10^{-6}$ of the extrapolation. With $r_d<r_f$ the premium at these spot levels is below $10^{-5}$, and the DSINC premium has an absolute error of the same order. This is the cancellation between the residue term and the spectral term described in \cref{rem:chichoice}.

\myparagraph{Correlated SABR.}
The last two rows of Table~\ref{tab:extra} price the American Puts of \cref{sec:sabrnum} with $\rho = \pm0.3$ in the coordinates \eqref{eq:sabrxy}, against the FD solver of that section on its coarse and fine grids. DSINC uses the frequency quadrature and the values of $N_y$, $N$ and $\chi$ of \cref{sec:numsetup}, on $\log y\in[\log y_0-4,\log y_0+3]$. At fixed $\sigma_0$, $y_0 = \sigma_0 f_0^{\beta-1}$ depends on the forward $f_0$, so the three spot levels are read from one run by cubic interpolation in $\log y$. The grid ends at $y\approx6$, where its Neumann condition replaces the vanishing killed solution.
At fixed volatility $\sigma_0 = 0.3$, this upper endpoint corresponds to a forward of approximately $0.0025$; at a fixed forward of one, it corresponds to a twentyfold increase in volatility. At this endpoint, the zero-order term of the fixed-frequency operator damps at a rate of at least $\tfrac12\chi(1-\chi)y^2\approx4.5$.

Moving the upper end to $\log y_0 + 4$ or $\log y_0 + 5$ at the same spacing, or replacing the Neumann condition by a zero value one step beyond the last node, changes the prices by less than $10^{-8}$. At $\rho = 0$ this DSINC reproduces the Hankel prices of Table~\ref{tab:sabr} to one unit in the sixth digit. With $\rho = \pm0.3$ it lies within $8\times10^{-6}$ of the fine grid and within $2.5\times10^{-6}$ of the extrapolation.

\begin{table}[!htb]
\centering
\caption{Further American Put tests, prices in units of the strike. The first two columns give the prices at $S/K = 1$. The last three columns give largest absolute values over $S/K\in\{0.9,1.0,1.1\}$. For SABR the forward $f$ takes the place of $S$. FD diff. is the difference between DSINC and the fine FD grid, FD change is the change of the FD prices under refinement, and Extrap. diff. is the difference between DSINC and the second-order extrapolation.}
\label{tab:extra}
\small
\setlength{\tabcolsep}{4pt}
\begin{tabular}{lccccc}
\toprule
Case & FD fine & DSINC & FD diff. & FD change & Extrap. diff. \\
\midrule
3/2, $T = 1$ & 0.063658 & 0.063668 & $9.7\times10^{-6}$ & $2.1\times10^{-5}$ & $2.6\times10^{-6}$ \\
Lognormal, $T = 1$ & 0.050718 & 0.050732 & $1.4\times10^{-5}$ & $4.0\times10^{-5}$ & $1.7\times10^{-6}$ \\
GARCH, $T = 1$ & 0.046411 & 0.046422 & $1.2\times10^{-5}$ & $3.2\times10^{-5}$ & $1.1\times10^{-6}$ \\
3/2, $r_d<r_f$ & 0.058611 & 0.058621 & $1.0\times10^{-5}$ & $2.7\times10^{-5}$ & $2.3\times10^{-6}$ \\
Lognormal, $r_d<r_f$ & 0.046948 & 0.046969 & $2.1\times10^{-5}$ & $3.9\times10^{-5}$ & $1.1\times10^{-5}$ \\
GARCH, $r_d<r_f$ & 0.045150 & 0.045168 & $1.8\times10^{-5}$ & $3.8\times10^{-5}$ & $7.4\times10^{-6}$ \\
3/2, strict local martingale & 0.049648 & 0.049662 & $1.4\times10^{-5}$ & $3.0\times10^{-5}$ & $4.0\times10^{-6}$ \\
GARCH, regime switch & 0.035568 & 0.035585 & $1.7\times10^{-5}$ & $4.2\times10^{-5}$ & $3.1\times10^{-6}$ \\
SABR, $\rho = -0.3$ & 0.082938 & 0.082946 & $8.0\times10^{-6}$ & $1.9\times10^{-5}$ & $2.5\times10^{-6}$ \\
SABR, $\rho = 0.3$ & 0.083222 & 0.083229 & $7.4\times10^{-6}$ & $1.9\times10^{-5}$ & $1.4\times10^{-6}$ \\
\bottomrule
\end{tabular}
\end{table}

\begin{table}[!htb]
\centering
\caption{Cost at equal estimated accuracy for the SABR Puts with \(\rho = \pm 0.3\), relative to the FD implementation used here. Elapsed time is reported in seconds, and the error is the largest estimated absolute error against the extrapolation from the fine and extra-fine grids, taken over both correlations and the three spot levels. For the extra-fine grid, the estimate is the size of the extrapolation correction.}
\label{tab:sabrcost}
\small
\sisetup{
    table-format=1.1e-1,
    round-mode=places,
    round-precision=1
}
\setlength{\tabcolsep}{4pt}
\begin{tabular}{@{}
    r @{\,} c @{\,} r    
    r                   
    S[table-format=2.1] 
    S                   
    @{\hspace{2em}}     
    r r                 
    S[table-format=2.1] 
    S                   
    @{}}
\toprule
\multicolumn{6}{c}{Finite differences} & \multicolumn{4}{c}{DSINC} \\
\cmidrule(r{2em}){1-6}\cmidrule(l){7-10}
\multicolumn{3}{c}{Nodes} & {Time steps} & {Time (s)} & {Est.\ error} &
\multicolumn{2}{c}{\(N_y\,\,\, N\)} & {Time (s)} & {Est.\ error} \\
\midrule
301  & \(\times\) & 41  & 150 & 0.2  & 2.6e-5 & 120 & 50  & 2.7  & 5.5e-6 \\
601  & \(\times\) & 81  & 300 & 2.1  & 6.5e-6 & 120 & 100 & 3.4  & 3.9e-6 \\
1201 & \(\times\) & 161 & 600 & 30.5 & 1.6e-6 & 160 & 100 & 8.8  & 2.6e-6 \\
     &            &     &     &      &        & 160 & 200 & 13.4 & 2.0e-6 \\
\bottomrule
\end{tabular}
\end{table}

Table~\ref{tab:sabrcost} compares the cost at equal estimated accuracy. The errors are estimated against the second-order extrapolation from the fine grid and an extra-fine grid with $1201\times161$ nodes and 600 steps, so the estimate for the extra-fine grid is the extrapolation correction itself. Relative to the FD implementation used here, which solves the full two-dimensional system at every step with one sparse LU factorization, the two methods cost about the same at the accuracy of the fine grid. That implementation is cheaper at lower accuracy, and DSINC is about twice as fast near $2\times10^{-6}$. A DSINC run has a fixed cost, the 108 matrix exponentials, which take about 2~s for $N_y = 120$ and 5~s for $N_y = 160$, plus a cost linear in $N$, and its error falls roughly in proportion to $\Delta t$. Halving every FD spacing divides the error by four but multiplies the time by 10 to 15. 

The relative cost also depends on the implementation of each method. ADI schemes, such as those of \cite{HaentjensInTHout2015}, and modern scheme as in \cite{Andersen2025}, could accelerate the FD reference, while Krylov approximations to the matrix-exponential actions, as discussed in \cite{ItkinDF2026}, could reduce the cost of DSINC, particularly on larger volatility grids. These alternatives have not been benchmarked here, so the reported timing ratios apply to the implementations tested.

\section{Conclusions} \label{sec:concl}

The collapse identity of DSINC is a statement about damped-sinc interpolation, and it holds for every one-factor SVM. Outside the affine class, what changes is the admissible damping and the evaluation of the volatility integral. The collapse identity retains its Fourier form for transition laws satisfying the stated integrability assumptions. The fundamental strip $0 < \chi < 1$ is available whenever the discounted spot is a nonnegative local martingale, with one residue term. Negative damping requires additional moments: it fails throughout the negative half-line for the GARCH and lognormal-volatility specifications considered here and for an absorbed SABR forward.

For the log-price-independent class \eqref{eq:gen}, the It\^o coordinate makes the endpoint phase explicit and supplies a natural coordinate for facets. In the 3/2 model, the reciprocal-CIR bridge gives the full-range facet transform through Kummer functions on an explicit joint moment domain, including the admissible critical tilt. The nodal fixed-frequency implementation integrates over terminal volatility directly and avoids the band-mass approximation. In SABR, the Bessel construction yields the premium integrand in closed form in the Hankel representation and independent volatility equations at zero correlation. With correlation, the volatility coordinate $\sigma f^{\beta-1}$ of \cite{Park2013} brings SABR into the class \eqref{eq:gen}, with a correlation that depends on the volatility and a volatility factor that explodes at absorption. The correlation expansion and the heat-kernel alternative retain the broader formulation developed in \cref{sec:sabr} and \cref{app:sabrheat}.

In the tests, DSINC with a nodal boundary prices American Puts in the 3/2, lognormal-volatility and GARCH models to within $1.8\times10^{-5}$ of a fine FD grid at a comparable cost, and its prices sit closer to the extrapolated FD values. The Hankel method prices American SABR Puts to within $8\times10^{-6}$ of the fine grid, but in our implementation it is slower than the two-dimensional solver. Eight further cases, with maturity one, $r_d<r_f$, a strict local martingale and a change of coefficients, keep DSINC within $2.1\times10^{-5}$ of the fine grid. For correlated SABR with $\rho = \pm0.3$, the fixed-frequency solution in the coordinates \eqref{eq:sabrxy} lies within $8\times10^{-6}$ of the fine grid, and at equal estimated accuracy it costs as much as our FD implementation near $6\times10^{-6}$ and half as much near $2\times10^{-6}$. For a prescribed boundary with one facet per date, the closed-form 3/2 premium agrees with the nodal one to $4\times10^{-7}$, but the single facet costs up to $1.8\times10^{-4}$ in price. The computed boundaries are closer to linear in the It\^o coordinate than in the variance, by factors of 1.4 to 4.5 in the largest residual. Table~\ref{tab:compare} summarizes what carries over from the Heston model.

\begin{table}[!htb]
\centering
\caption{What carries over from the Heston model.}
\label{tab:compare}
\small
\begin{tabular}{>{\raggedright\arraybackslash}p{0.16\textwidth}>{\raggedright\arraybackslash}p{0.16\textwidth}>{\raggedright\arraybackslash}p{0.18\textwidth}>{\raggedright\arraybackslash}p{0.17\textwidth}>{\raggedright\arraybackslash}p{0.17\textwidth}}
\toprule
Component & Heston & 3/2 model & SABR, $\beta<1$ & Model \eqref{eq:gen} \\
\midrule
Collapse identity and Parseval reduction & Hold & Hold unchanged & Hold unchanged & Hold unchanged \\
Admissible damping & Negative or strip & Above the bound \eqref{eq:adm32}, or strip & Strip only & Strip, negative model specific \\
Endpoint contribution to the conditional log-price mean & Linear in $v_u$ & Linear in $\log v_u$ & Not applicable, the log-price is not conditionally Gaussian & Linear in $F(y_u)$ in \eqref{eq:gen} \\
Bridge factor given the endpoints & Tilted CIR kernel divided by the CIR transition density & Ratio of Bessel functions with a shifted order & Hankel representation for $\rho = 0$ & Bridge expectation in \eqref{eq:bridgefun} \\
Outer volatility integral, full range & Affine transform and M\"obius propagator for facets linear in $v$ & Kummer transform for facets linear in $\log v$ & Premium integrand in closed form in the Hankel representation, independent volatility equations at $\rho = 0$. Fixed-frequency solution in $(x,y)$ for any $\rho$ & Fixed-frequency solution for \eqref{eq:gen}, nodal boundary \\
Deterministic $\theta(t)$ & Affine Riccati system & Closed form after a time change & Not in the model & Time-dependent operator \\
Piecewise-constant parameters & Exact shear at each breakpoint & One quadrature per breakpoint crossed & Operator switch & Operator switch \\
\bottomrule
\end{tabular}
\end{table}

The conclusion of \cite{AndersenItkinKazbek2026} can therefore be restated so that it holds beyond the Heston model.
\begin{quote}
\begin{quote}
For any one-factor SVM satisfying the stated assumptions, the sum over the local basis reduces exactly to the characteristic function conditional on the terminal volatility, evaluated on a reflected contour, so the log-price integration becomes a single spectral integral. For a full-range facet, the remaining volatility integral is available in closed form in the Heston model when the exercise boundary is linear in the variance, and in the 3/2 model when it is linear in the logarithm of the variance. For models whose coefficients are independent of the log-price, possibly after a change of variables, the remaining volatility integral can otherwise be evaluated by one backward equation per frequency, which also makes facets unnecessary.
\end{quote}
\end{quote}

Several extensions are natural. The 4/2 model of \cite{Grasselli2017} has an It\^o coordinate that mixes the Heston and 3/2 cases. For SABR, the correlation expansion of \cref{sec:sabrcorr} remains to be implemented and tested. Flexible forwards can be treated in the same way where the Put--Call duality holds, which requires a true martingale. For coefficients independent of the log-price, two volatility factors turn the fixed-frequency problem into a two-dimensional one while still avoiding a grid in the log-price.

\section*{Disclosure statement}
No potential conflict of interest was reported by the author.

\section*{Funding}
No funding was received.

\section*{Disclaimer}
Opinions expressed here are the author's own, and do not represent the views of their employers. A standard disclaimer applies.

\section*{Acknowledgments}

I thank Leif Andersen and Rakhymzhan Kazbek for our joint work on the previous paper, which stimulated the additional ideas explored in this work.

\medskip
\noindent During the preparation of this work, the author used GPT-6 Astra and Claude Opus 5.5 for the purpose of validating mathematical proofs, reviewing Python source code, and refining the English language and phrasing of the manuscript. Following the use of these tools, the author thoroughly reviewed, verified, and edited the output. The author assumes full responsibility for the mathematical accuracy, logical integrity of the proofs, and final content of this manuscript.

\printbibliography[title = {References}]

@Article{Andersen2025,
  title     = {American Option Pricing by the Finite Difference Method},
  author    = {L. Andersen},
  year      = {2026},
  month     = mar,
  journal   = {Wilmott},
  volume    = {2026},
  number    = {142},
  issn      = {1541-8286},
  doi       = {10.54946/wilm.12224},
  publisher = {Wilmott Magazine Ltd},
}

@article{BroadieKaya2006,
  author    = {Broadie, Mark and Kaya, \"{O}mer},
  title     = {Exact Simulation of Stochastic Volatility and Other Affine Jump Diffusion Processes},
  journal   = {Operations Research},
  year      = {2006},
  volume    = {54},
  number    = {2},
  pages     = {217--231},
  doi       = {10.1287/opre.1050.0247}
}

@TechReport{Chiarella2004,
  author      = {C. Chiarella and A. Kucera and A. Ziogas},
  institution = {Quantitative Finance Research Centre, University of Technology, Sydney},
  title       = {A survey of the integral representation of {American} option prices},
  year        = {2004},
  number      = {118},
}

@InProceedings{Chiarella2005,
  author       = {C. Chiarella and A. Ziogas},
  title        = {Pricing {American} Options under Stochastic Volatility},
  booktitle    = {Computing in Economics and Finance 2005},
  number       = {77},
  
  publisher    = {Society for Computational Economics},
  year         = {2005},
  url          = {https://ideas.repec.org/p/sce/scecf5/77.html}
}

@Article{Chiarella2009,
  author       = {C. Chiarella and A. Ziogas},
  journal      = {Applied Mathematical Finance},
  title        = {American Call Options Under Jump-Diffusion Processes -- A Fourier Transform Approach},
  year         = {2009},
  number       = {1},
  pages        = {37--79},
  volume       = {16},
  date         = {2009},
  journaltitle = {Applied Mathematical Finance},
}

@Article{DetempleKitapbayev2017,
  author  = {J. Detemple and Y. Kitapbayev},
  journal = {Mathematical Finance},
  title   = {On {American VIX} options under the generalized 3/2 and 1/2 models},
  year    = {2018},
  volume  = {28},
  number  = {2},
  pages   = {550--581},
  doi     = {10.1111/mafi.12153},
}

@Article{Drimus2012,
  author  = {G.G. Drimus},
  journal = {Quantitative Finance},
  title   = {Options on realized variance by transform methods: a non-affine stochastic volatility model},
  year    = {2012},
  number  = {11},
  pages   = {1679--1694},
  volume  = {12},
}

@Article{Itkin2024jd,
  author = {Andrey Itkin},
  year   = {2024},
  title  = {Semi-analytic pricing of {American} options in time-dependent jump-diffusion models with exponential jumps},
  journal = {The Journal of Derivatives},
  volume = 31, 
  number  = {4},
}

@Article{ItkinCF2025j,
  author       = {Andrey Itkin},
  date         = {2026},
  journaltitle = {Review of Derivatives Research},
  title        = {American options valuation in time-dependent jump-diffusion models via integral equations and characteristic functions},
  number       = {10},
  volume       = {29},
  year         = {2026},
}

@Article{ItkinKitapbayev2025r,
  author = {Andrey Itkin and Yerkin Kitapbayev},
  journal = {Risk},
  month  = dec,
  title  = {Floating exercise boundaries for {American} options in time-inhomogeneous models},
  year   = {2025},
}

@Article{ItkinMuravey2024jd,
  author  = {Andrey Itkin and Dmitry Muravey},
  year    = {2024},
  title   = {American options in time-dependent one-factor models: Semi-analytic pricing, numerical methods, and {ML} support},
  journal = {The Journal of Derivatives},
  volume  = {31},
  number  = {3},
  pages   = {74--114},
  doi     = {10.3905/jod.2023.1.197},
}

@Article{Peskir2005,
  author  = {G. Peskir},
  journal = {Journal of Theoretical Probability},
  title   = {A Change-of-Variable Formula with Local Time on Curves},
  year    = {2005},
  volume  = {18},
  number  = {3},
  pages   = {499--535},
  doi     = {10.1007/s10959-005-3517-6}
}

@incollection{Peskir2007,
  author    = {Peskir, Goran},
  title     = {A Change-of-Variable Formula with Local Time on Surfaces},
  booktitle = {S{\'e}minaire de Probabilit{\'e}s {XL}},
  editor    = {Donati-Martin, Catherine and {\'E}mery, Michel and Rouault, Alain and Stricker, Christophe},
  series    = {Lecture Notes in Mathematics},
  volume    = {1899},
  pages     = {69--96},
  publisher = {Springer},
  address   = {Berlin, Heidelberg},
  year      = {2007},
  doi       = {10.1007/978-3-540-71189-6_2}
}

@techreport{tzavalis2003pricing,
  author      = {Tzavalis, Elias and Wang, Shijun},
  title       = {Pricing {American} Options under Stochastic Volatility: A New Method Using {Chebyshev} Polynomials to Approximate the Early Exercise Boundary},
  institution = {Department of Economics, Queen Mary, University of London},
  type        = {Working Paper},
  number      = {488},
  year        = {2003},
  month       = feb,
  url         = {https://www.qmul.ac.uk/sef/media/econ/research/workingpapers/2003/items/wp488.pdf}
}

@Article{CarrSun2007,
  Title                    = {A New Approach for Option Pricing Under Stochastic Volatility},
  Author                   = {Peter Carr and Jian Sun},
  Journal                  = {Review of Derivatives Research},
  Year                     = {2007},
  Pages                    = {87-250},
  Volume                   = {10},
}

@article{FangOosterlee2011,
  author = {Fang, Fang and Oosterlee, Cornelis W.},
  title = {A Fourier-Based Valuation Method for Bermudan and Barrier Options under Heston's Model},
  journal = {SIAM Journal on Financial Mathematics},
  volume = {2},
  number = {1},
  pages = {439--463},
  year = {2011},
  publisher = {Society for Industrial and Applied Mathematics},
}

@article{ItkinDF2026,
  title		={Diagonal Frog: High-order positivity-preserving FD schemes for anisotropic Fokker-Planck equations},
  author	={Itkin, Andrey},
  year		={2026},
  eprint	={2606.23980},
  archivePrefix	={arXiv},
  primaryClass	={math.NA},
}

@book{Lewis2000,
  author    = {Lewis, Alan L.},
  title     = {Option Valuation under Stochastic Volatility: With Mathematica Code},
  publisher = {Finance Press},
  address   = {Newport Beach, CA},
  year      = {2000}
}

@misc{Lewis2001,
  author       = {Lewis, Alan L.},
  title        = {A simple option formula for general jump-diffusion and other exponential {L}\'evy processes},
  howpublished = {Working paper, OptionCity.net},
  year         = {2001}
}

@article{BaldeauxBadran2014,
  author  = {Baldeaux, Jan and Badran, Alexander},
  title   = {Consistent modelling of {VIX} and equity derivatives using a 3/2 plus jumps model},
  journal = {Applied Mathematical Finance},
  volume  = {21},
  number  = {4},
  pages   = {299--312},
  year    = {2014}
}

@article{ChristoffersenJacobsMimouni2010,
  author  = {Christoffersen, Peter and Jacobs, Kris and Mimouni, Karim},
  title   = {Volatility dynamics for the {S\&P500}: evidence from realized volatility, daily returns, and option prices},
  journal = {Review of Financial Studies},
  volume  = {23},
  number  = {8},
  pages   = {3141--3189},
  year    = {2010}
}

@article{KaeckAlexander2012,
  author  = {Kaeck, Andreas and Alexander, Carol},
  title   = {Volatility dynamics for the {S\&P} 500: further evidence from non-affine, multi-factor jump diffusions},
  journal = {Journal of Banking \& Finance},
  volume  = {36},
  number  = {11},
  pages   = {3110--3121},
  year    = {2012}
}

@article{Grasselli2017,
  author  = {Grasselli, Martino},
  title   = {The 4/2 stochastic volatility model: a unified approach for the {H}eston and the 3/2 model},
  journal = {Mathematical Finance},
  volume  = {27},
  number  = {4},
  pages   = {1013--1034},
  year    = {2017}
}

@article{Nelson1990,
  author  = {Nelson, Daniel B.},
  title   = {{ARCH} models as diffusion approximations},
  journal = {Journal of Econometrics},
  volume  = {45},
  number  = {1--2},
  pages   = {7--38},
  year    = {1990}
}

@article{BaroneAdesiEtAl2005,
  author  = {Barone-Adesi, Giovanni and Rasmussen, Henrik and Ravanelli, Claudia},
  title   = {An option pricing formula for the {GARCH} diffusion model},
  journal = {Computational Statistics \& Data Analysis},
  volume  = {49},
  number  = {2},
  pages   = {287--310},
  year    = {2005}
}

@article{Hagan2002,
  author  = {Hagan, Patrick S. and Kumar, Deep and Lesniewski, Andrew S. and Woodward, Diana E.},
  title   = {Managing smile risk},
  journal = {Wilmott Magazine},
  month   = sep,
  pages   = {84--108},
  year    = {2002}
}

@article{Itkin2013,
  author  = {Itkin, Andrey},
  title   = {New solvable stochastic volatility models for pricing volatility derivatives},
  journal = {Review of Derivatives Research},
  volume  = {16},
  number  = {2},
  pages   = {111--134},
  year    = {2013}
}

@article{ClarkeParrott1999,
  author  = {Clarke, Nigel and Parrott, Kevin},
  title   = {Multigrid for {A}merican option pricing with stochastic volatility},
  journal = {Applied Mathematical Finance},
  volume  = {6},
  number  = {3},
  pages   = {177--195},
  year    = {1999}
}

@article{IkonenToivanen2004,
  author  = {Ikonen, Samuli and Toivanen, Jari},
  title   = {Operator splitting methods for {A}merican option pricing},
  journal = {Applied Mathematics Letters},
  volume  = {17},
  number  = {7},
  pages   = {809--814},
  year    = {2004}
}

@article{IkonenToivanen2009,
  author  = {Ikonen, Samuli and Toivanen, Jari},
  title   = {Operator splitting methods for pricing {A}merican options under stochastic volatility},
  journal = {Numerische Mathematik},
  volume  = {113},
  number  = {2},
  pages   = {299--324},
  year    = {2009}
}

@article{HaentjensInTHout2015,
  author  = {Haentjens, Tinne and {in 't Hout}, Karel J.},
  title   = {{ADI} schemes for pricing {A}merican options under the {H}eston model},
  journal = {Applied Mathematical Finance},
  volume  = {22},
  number  = {3},
  pages   = {207--237},
  year    = {2015}
}

@article{MareeOrtizGraciaOosterlee2017,
  author  = {Maree, Stefanus C. and Ortiz-Gracia, Luis and Oosterlee, Cornelis W.},
  title   = {Pricing early-exercise and discrete barrier options by {S}hannon wavelet expansions},
  journal = {Numerische Mathematik},
  volume  = {136},
  number  = {4},
  pages   = {1035--1070},
  year    = {2017}
}

@article{BaschettiEtAl2022,
  author  = {Baschetti, Fabio and Bormetti, Giacomo and Romagnoli, Silvia and Rossi, Pietro},
  title   = {The {SINC} way: a fast and accurate approach to {F}ourier pricing},
  journal = {Quantitative Finance},
  volume  = {22},
  number  = {3},
  pages   = {427--446},
  year    = {2022}
}

@article{AndersenPiterbarg2007,
  author  = {Andersen, Leif B. G. and Piterbarg, Vladimir V.},
  title   = {Moment explosions in stochastic volatility models},
  journal = {Finance and Stochastics},
  volume  = {11},
  number  = {1},
  pages   = {29--50},
  year    = {2007}
}

@article{LionsMusiela2007,
  author  = {Lions, Pierre-Louis and Musiela, Marek},
  title   = {Correlations and bounds for stochastic volatility models},
  journal = {Annales de l'Institut Henri Poincar\'e C, Analyse non lin\'eaire},
  volume  = {24},
  number  = {1},
  pages   = {1--16},
  year    = {2007}
}

@techreport{Jourdain2004,
  author      = {Jourdain, Benjamin},
  title       = {Loss of martingality in asset price models with lognormal stochastic volatility},
  institution = {CERMICS},
  type        = {Preprint},
  number      = {2004-267},
  year        = {2004}
}

@article{HullWhite1987,
  author  = {Hull, John and White, Alan},
  title   = {The pricing of options on assets with stochastic volatilities},
  journal = {Journal of Finance},
  volume  = {42},
  number  = {2},
  pages   = {281--300},
  year    = {1987}
}

@article{Yor1992,
  author  = {Yor, Marc},
  title   = {On some exponential functionals of {B}rownian motion},
  journal = {Advances in Applied Probability},
  volume  = {24},
  number  = {3},
  pages   = {509--531},
  year    = {1992}
}

@article{Higham2005,
  author  = {Higham, Nicholas J.},
  title   = {The scaling and squaring method for the matrix exponential revisited},
  journal = {SIAM Journal on Matrix Analysis and Applications},
  volume  = {26},
  number  = {4},
  pages   = {1179--1193},
  year    = {2005}
}

@article{MolerVanLoan2003,
  author  = {Moler, Cleve and Van Loan, Charles},
  title   = {Nineteen dubious ways to compute the exponential of a matrix, twenty-five years later},
  journal = {SIAM Review},
  volume  = {45},
  number  = {1},
  pages   = {3--49},
  year    = {2003}
}

@article{Willard1997,
  author  = {Willard, Gregory A.},
  title   = {Calculating prices and sensitivities for path-independent derivative securities in multifactor models},
  journal = {Journal of Derivatives},
  volume  = {5},
  number  = {1},
  pages   = {45--61},
  year    = {1997}
}

@article{Rannacher1984,
  author  = {Rannacher, Rolf},
  title   = {Finite element solution of diffusion problems with irregular data},
  journal = {Numerische Mathematik},
  volume  = {43},
  number  = {2},
  pages   = {309--327},
  year    = {1984}
}

@article{ItkinMuravey2021SABR,
  author    = {Itkin, Andrey and Muravey, Dmitry},
  title     = {Semi-Analytical Pricing of Barrier Options in the Time-Dependent $\lambda$-{SABR} Model: {Uncorrelated} Case},
  journal   = {The Journal of Derivatives},
  volume    = {30},
  number    = {1},
  pages     = {74--101},
  year      = {2022},
  publisher = {Pageant Media},
}

@misc{Paulot2009,
  author       = {Paulot, Louis},
  title        = {Asymptotic implied volatility at the second order with application to the {SABR} model},
  howpublished = {arXiv preprint 0906.0658},
  year         = {2009}
}

@article{AntonovKonikovSpector2013,
  author  = {Antonov, Alexandre and Konikov, Michael and Spector, Michael},
  title   = {{SABR} spreads its wings},
  journal = {Risk},
  month   = aug,
  year    = {2013}
}

@book{NIST2010,
  author    = {Olver, Frank W. J. and Lozier, Daniel W. and Boisvert, Ronald F. and Clark, Charles W.},
  title     = {{NIST} Handbook of Mathematical Functions},
  publisher = {Cambridge University Press},
  address   = {Cambridge},
  year      = {2010}
}

@misc{AndersenItkinKazbek2026,
  author        = {Andersen, Leif and Itkin, Andrey and Kazbek, Rakhymzhan},
  title         = {Valuing American options and Flexible Forwards contracts in time-dependent models},
  year          = {2026},
  eprint        = {2606.27335},
  archivePrefix = {arXiv},
  primaryClass  = {q-fin.PR}
}

@article{PitmanYor1982,
  author    = {Pitman, Jim and Yor, Marc},
  title     = {A decomposition of Bessel bridges},
  journal   = {Zeitschrift f{\"u}r Wahrscheinlichkeitstheorie und Verwandte Gebiete},
  volume    = {59},
  number    = {4},
  pages     = {425--457},
  year      = {1982},
  publisher = {Springer},
}

@Book{Antonov2019,
  author    = {A. Antonov and M. Konikov and M. Spector},
  year      = {2019},
  title     = {Modern SABR Analytics},
  isbn      = {978-3-030-10656-0},
  publisher = {Springer},
  series    = {Springer Briefs in Quantitative Finance},
}

@article{Baldeaux2012ExactSimulation,
  author = {Baldeaux, Jan},
  title = {Exact Simulation of the {3/2} Model},
  journal = {International Journal of Theoretical and Applied Finance},
  year = {2012},
  volume = {15},
  number = {5},
  pages = {1250032},
}

@article{ZhengZeng2016PartialTransform,
  author = {Zheng, Wendong and Zeng, Pingping},
  title = {Pricing timer options and variance derivatives with closed-form partial transform under the {3/2} model},
  journal = {Applied Mathematical Finance},
  year = {2016},
  volume = {23},
  number = {5},
  pages = {344--373},
}

@article{CrankNicolson1947, 
    author  = {Crank, J. and Nicolson, P.}, 
    title   = {A practical method for numerical evaluation of solutions of partial differential equations of the heat-conduction type}, 
    journal = {Mathematical Proceedings of the Cambridge Philosophical Society}, 
    volume  = {43}, 
    number  = {1}, 
    pages   = {50--67}, 
    year    = {1947}
}

@article{inTHoutFoulon2010, 
    author  = {{in 't Hout}, K. J. and Foulon, S.}, 
    title   = {{ADI} finite difference schemes for option pricing in the {H}eston model with correlation},
    journal = {International Journal of Numerical Analysis and Modeling}, 
    volume  = {7}, 
    number  = {2}, 
    pages   = {303--320}, 
    year    = {2010}
}

@misc{Park2013,
    author       = {Park, H.},
    title        = {Efficient valuation method for the {SABR} model},
    year         = {2013},
    eprint       = {1308.0665},
    archivePrefix = {arXiv},
    primaryClass = {q-fin.CP}
}

\appendix
\label{app:proofs}
\appendixpage
\numberwithin{equation}{section}
\setcounter{equation}{0}

\section{Duhamel iteration for correlated SABR} \label{app:corr}

Let $\mathcal{L}_{0,t}$ be the generator \eqref{eq:sabrgen} at $\rho = 0$, and let
\begin{equation} \label{eq:sabrD}
\mathcal{D}_t = \bar\rho(t)\,\varepsilon(t)\,\sigma^2\,\partial_{R\sigma},
\end{equation}
so that $\mathcal{L}_t = \mathcal{L}_{0,t} + c\,\mathcal{D}_t$. For a terminal datum $H$ on the positive-price half-line, the transition semigroup of $\mathcal{L}_t$ satisfies the Duhamel identity
\begin{equation} \label{eq:sabrduhamel}
w(t) = P_0(t,u)H + \int_t^u P_0(t,s)\,c\,\mathcal{D}_s\,w(s)\,ds,
\end{equation}
where $P_0(t,s)$ is the $\rho = 0$ propagator.

By \eqref{eq:sabrkernel}, $P_0(t,s)$ acts on a function $g(R,\sigma)$ as
\begin{equation} \label{eq:sabrP0}
\bigl[P_0(t,s)g\bigr](R,\sigma) = \int_0^\infty q\,\Phi_q(R)\,
\widehat G_{\hat g(q,\cdot)}(t,\sigma ; s,q)\,dq,
\end{equation}
with $\hat g(q,\sigma) = \int_0^\infty R_u^{1-2 \nu_\beta} \Phi_q(R_u)\,g(R_u, \sigma)\,dR_u$ the weighted Hankel transform. The Neumann series is $w = \sum_{n\ge0}c^nw_n$ with
\begin{equation} \label{eq:sabrneumann}
w_0 = P_0(t,u)H, \qquad
w_n(t) = \int_t^u P_0(t,s)\,\bar\rho(s)\varepsilon(s)\sigma^2\,
\partial_{R\sigma}w_{n-1}(s)\,ds, \qquad n\ge1 .
\end{equation}
The series \eqref{eq:sabrneumann} is \eqref{eq:sabrpert} written in integral form. Its partial sum up to order $N_h$ coincides with the $N_h$-th Picard iterate of \eqref{eq:sabrduhamel} started from $w_0$, so iterating the Duhamel map and truncating the expansion are the same approximation.

The mixed derivative acts on $\Phi_q\widehat G$ as $\partial_R\Phi_q\,\partial_\sigma\widehat G + \Phi_q\,\partial_{R\sigma}\widehat G$. By \cref{rem:coupling}, the factor $\partial_R\Phi_q = qR^{\nu_\beta}J_{\nu_\beta-1}(qR)$ is not a multiple of any single $\Phi_{q'}$, so the source of each order does not stay within one Bessel frequency. Each order requires one volatility equation per Bessel frequency, with the Hankel transform of the new source recomputed, rather than one independent equation per frequency. Near $R = 0$ the regular solution behaves like $R^{2\nu_\beta}$, so $\partial_R$ produces a factor $R^{2\nu_\beta-1}$, which vanishes there because $\nu_\beta>1/2$.

Convergence of \eqref{eq:sabrneumann} at the target correlation is not established here. The operator $\mathcal{D}_s$ is of second order and carries the unbounded factor $\sigma^2$. A convergence proof would need smoothing estimates for $P_0$ that control both derivatives, which we do not provide. The number of orders needed for a given accuracy must therefore be determined numerically.

The iteration is compatible with the boundary computation of \cref{sec:sabramer}. The exercise boundary enters only through the terminal datum $H = \widehat P(q,\cdot)$ of \eqref{eq:sabrPhat}, so the orders can be nested inside the backward recursion \eqref{eq:sabrQ}. The cost of the correlated solution is then the uncorrelated cost times the number of orders, plus one Hankel transform of the source per order and frequency. Since the uncorrelated Bessel implementation is already slower than finite FD in \cref{sec:sabrnum}, we do not claim a speed advantage for the correlated iteration.

\section{A heat-kernel alternative for correlated SABR} \label{app:sabrheat}

For classical SABR, with $\kappa = 0$ and constant $\varepsilon$ and $\rho$, a further approximation follows from the joint heat-kernel expansion that underlies \cite{AntonovKonikovSpector2013, Antonov2019}. Its small parameter is the time lag $\tau = u-t$, while $\rho\in(-1,1)$ enters the coefficients without an expansion about zero. The joint expansion is developed in \cite{Paulot2009}. The terminal volatility must be retained for the American exercise problem. An interior approximation of the joint density with respect to $df_u\,d\sigma_u$ has the form
\begin{equation} \label{eq:sabrheat}
p^{(N_h)}(\tau ; f,\sigma;f_u,\sigma_u) = \frac{\mathcal{H}(f,\sigma ; f_u,\sigma_u)}{2\pi\tau} \exp\Bigl(-\frac{d(f,\sigma ; f_u,\sigma_u)^2}{2\tau}\Bigr)
\sum_{n = 0}^{N_h} a_n(f,\sigma ; f_u,\sigma_u)\, \tau^n, \qquad a_0  =  1.
\end{equation}
Here $d$ is the geodesic distance for the inverse covariance metric, $\mathcal{H}$ collects the density Jacobian, the geometric determinant and the parallel-transport factor, and $N_h$ is the truncation order. The coefficients $a_n$ are specified by the heat-kernel construction. The sum in \eqref{eq:sabrheat} is an asymptotic truncation, not a convergent series for arbitrary $\tau$.

For an approximation $\widehat p^{\, + }$ of the positive-price density with the required weighted integrability, the DSINC input is
\begin{equation} \label{eq:sabrheatpsi}
\widehat\Psi(\widetilde\omega,\sigma_u ; x,\sigma)
 = \int_0^\infty\Bigl(\frac{f_u}{K}\Bigr)^{\chi-\iu\omega}
\widehat p^{\, + }(\tau ; f,\sigma ; f_u,\sigma_u)\,df_u .
\end{equation}
The spectral collapse then applies to this approximation under the assumptions of \cref{prop:sabrpairing}, and its error enters \eqref{eq:sabrerror}. This approach requires a numerical transformation and does not give independent one-dimensional evolution equations for correlated SABR.

Two qualifications govern its use. The interior expansion does not enforce absorption at $f = 0$, \cite{Paulot2009}. The payoff subtraction already handles the atom, but accuracy of the continuous density near zero still requires boundary treatment or validation. Moreover, \eqref{eq:sabrprice} uses all lags $u-t$, so refinement of the exercise-time grid alone does not justify a short-time approximation. Longer lags require a validated kernel approximation or a composition of suitably controlled short-step transition operators. The strike-dependent European-price mapping of \cite{AntonovKonikovSpector2013, Antonov2019} does not by itself supply this joint operator. We therefore retain the Bessel construction as the principal development at zero correlation, with the Duhamel iteration of \cref{app:corr} as its correlated extension. For correlated SABR, the implemented method is the fixed-frequency solution in the coordinates \eqref{eq:sabrxy} of \cref{sec:sabrcorr}, and the heat-kernel approach remains an additional approximation, to be assessed against a two-dimensional solver.

\end{document}